%% file: text7.tex
\documentclass[traditabstract]{aa}
\usepackage{natbib}
\usepackage{graphicx}
\usepackage{txfonts}
\usepackage{xspace}
\usepackage{linenoaa}
\begin{document} 
\title{Chemical composition and kinematics of ionised gas in low-mass
  star-forming galaxies with extremely high [O~{\sc iii}]/[O~{\sc ii}] ratios\thanks{Based on observations collected at the European Southern Observatory under ESO programme
0105.B-0772(A)}}
\author{Y. I. Izotov
          \inst{1}
          \and
          N. G. Guseva
          \inst{1}
          \and
          D. Schaerer
          \inst{2,3}
         \and
         R. O. Amor\'in
         \inst{4}
           }

   \institute{ Bogolyubov Institute for Theoretical Physics,
                     National Academy of Sciences of Ukraine, 
                     14-b Metrolohichna str., Kyiv, 03143, Ukraine\\
              \email{yizotov@bitp.kyiv.ua}
         \and
            Department of Astronony, University of Geneva, 
                     51 Ch. des Maillettes, 1290, Versoix, Switzerland
\and
IRAP/CNRS, 14, Av. E. Belin, 31400 Toulouse, France
\and
Instituto de Astrof\'{i}sica de Andaluc\'{i}a (CSIC), Apartado 3004, 18080
Granada, Spain
             }

\date{Received April 15, 2023; accepted ?????????}

\abstract{
We present Very Large Telescope/Xshooter spectrophotometric observations
of eleven low-redshift ($z$\,$<$\,0.085) compact star-forming galaxies
(`high O$_{32}$ sample').
These galaxies are characterized by extremely high emission-line ratios
[O~{\sc iii}]$\lambda$5007/[O~{\sc ii}]$\lambda$3727, ranging from
11 to 42. Galaxies with such high ratios are thought to be promising 
candidates for leaking large amounts of Lyman continuum radiation.
They are characterized by low oxygen abundances 12\,+\,log(O/H)\,=\,$7.5-8.0$ 
and low stellar masses $M_\star$\,$\sim$\,$10^6-10^8$ M$_\odot$.
Strong emission lines of various ions in all spectra are
used to derive helium and oxygen abundances,
and N/O, Ne/O, S/O, Cl/O, Ar/O and Fe/O abundance ratios.
We also derived macroscopic velocity dispersions
$\sigma$($\lambda$) from various emission lines of different ions. We find that 
$\sigma$(4861) of the H$\beta$ emission line is increased with increasing
stellar mass and decreasing O$_{32}$ ratio. On the other hand, the
$\sigma$($\lambda$)/$\sigma$(4861) ratios for various lines are close to 1.
Exceptions are $\sigma$($\lambda$)/$\sigma$(4861) of two lines,
He~{\sc ii} 4686 and He~{\sc i} 10830, which are considerably higher than unity
and of four lines, [O~{\sc ii}] 3726,  3729, [S~{\sc ii}] 6717, 6731, with
$\sigma$($\lambda$)/$\sigma$(4861) lower than unity. The two former lines are
likely produced in the inner parts of H~{\sc ii} regions and are
broadened by dynamical processes generated by massive stars, and by
radiative scattering in the case of the He~{\sc i} 10830 emission line.
Emission in the four latter lines is produced mainly in the outer and likely
more quiet parts of H~{\sc ii} regions.}

\titlerunning{Galaxies with extremely high [O~{\sc iii}]/[O~{\sc ii}] ratios}
   \maketitle
   \nolinenumbers
   
   \section{Introduction}

The studies of low-redshift compact star-forming galaxies (CSFGs) selected
from the Sloan Digital Sky Survey (SDSS) have shown that they are low-mass and
low-metallicity galaxies with star formation occuring in short bursts of 
a few million years duration \citep{I06,I11,I18c,I21a,A10}.
 
\input{tab1.tex}

Their properties, together with the fact that CSFGs and high-redshift 
star-forming galaxies follow similar mass-metallicity and 
luminosity-metallicity relations, suggest that CSFGs are good local
counterparts of the high-redshift dwarf star-forming galaxies (SFGs)
\citep[e.g. ][]{I21a}. The
strong emission lines in the optical spectra of H~{\sc ii} regions in CSFGs
are powered by numerous O-stars, which produce plenty of 
ionising radiation. This makes CSFGs promising candidates for considerable
leaking ionising radiation into the intergalactic medium (IGM). 
It is indeed thought that dwarf SFGs with similar properties are responsible
for the reionisation of the Universe at redshifts $z$ = 5 -- 10 
\citep[e.g. ][]{O09,M13,Y11,B15,At24}. 

Many CSFGs are characterised by high line ratios O$_{32}$ = 
[O~{\sc iii}]$\lambda$5007/[O~{\sc ii}]$\lambda$3727, reaching values of up to
60 in some galaxies \citep[e.g. ][]{S15,I21b,I24}. Such high values
may indicate that their H~{\sc ii} regions are density-bounded, allowing escape
of ionising radiation into the IGM, as suggested e.g.\ by \citet{JO13}, 
\citet{NO14} and \citet{N16}. However, that is not the 
only explanation. High O$_{32}$
can also be caused by a low metallicity, a high ionisation  parameter
or a hard  ionising  radiation \citep{JO13,S15}. Unfortunately, direct
measurenents of Lyman continuum (LyC) emission are not possible for most of
CSFGs because of their low redshifts, including galaxies, considered in this
paper. This fact does not allow to link high O$_{32}$ with either of above
mechanisms.

The properties of the vast majority of CSFGs
derived in the optical range were based on low-resolution spectra with
unresolved nebular emission lines. On the other
hand, higher resolution spectroscopic data allow to study kinematic properties
of the ionised gas in CSFGs. However, until now mainly
some strongest lines, most commonly only H$\alpha$, [O~{\sc iii}] $\lambda$5007,
H$\beta$, were considered \citep[e.g. ][]{BT11,Ch14,M17,A12,A24}.

In this paper, we have used Xshooter spectrograph mounted on the Very Large
Telescope (VLT) to obtain spectroscopic observations with high signal-to-noise
ratio of 11 CSFGs at $z$~$<$~0.085 with
extremely high O$_{32}$ = 11 -- 42 to derive their chemical composition and
to study kinematics of the ionised gas using various strong emission lines of
ions of different elements and with different ionisation potentials. 
In Section \ref{sec:selection} we describe the selection criteria of CFSGs and
their global characteristics. The observations and data reduction are described
in Section \ref{sec:observations}. Element abundances are derived in Section
\ref{sec:abundances}. In Section \ref{sec:kinematics} we study kinematics of the
ionised gas.
We summarise our main results in Section 
\ref{sec:conclusions}.

\section{Selection of CSFGs and their global characteristics}\label{sec:selection}

The objects were selected from the SDSS by their
compactness, strong emission lines, extremely high emission-line ratios
O$_{32}$~=~[O~{\sc iii}]$\lambda$5007/[O~{\sc ii}]$\lambda$3727 and absence of
Active Galactic Nuclei (AGNs) spectral features. Additionally, the objects must
have low declinations $\la$ 20 degree to be
accessible for observations with the VLT.
The sample includes 11 relatively bright galaxies at $z$ $<$ 0.085
(Table \ref{tab1}). Some of these galaxies have been studied in the
UV range with the {\sl Hubble Space Telescope} ({\sl HST})
\citep[emission in Ly$\alpha$ and other UV lines, e.g. ][]{J17,I20,Be22,Xu22}, 
in the optical range \citep[e.g. ][]{I11,I17,Ch14,Y17,K24},
and in the radio range \citep{Fi13}. Their angular sizes on the images obtained with the {\sl HST} are considerably lower than those measured from the SDSS
images (Table~\ref{tab2}).

\begin{figure*}[t]
\hbox{
\includegraphics[angle=0,width=0.99\linewidth]{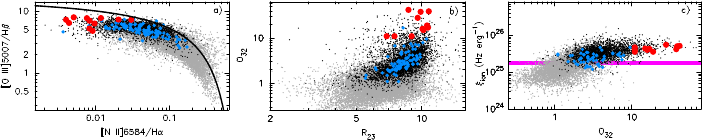}
}
\caption{a) BPT diagram \citep{BPT81}.
Galaxies from this paper and from
\citet{Ch14} are shown by red circles and blue circles, respectively.
Compact SFGs from the SDSS
are represented by black dots (EW(H$\beta$)~$\geq$~100\AA) and grey
dots (EW(H$\beta$)~$<$~100\AA). The line separating SFGs and AGNs is from
\citet{K03}. b) O$_{32}$ -- R$_{23}$ diagram, where
O$_{32}$~=~$I$([O~{\sc iii}] $\lambda$5007)/$I$([O~{\sc ii}] $\lambda$3727) and
R$_{23}$ = $I$([O~{\sc ii}] $\lambda$3727 + [O~{\sc iii}] $\lambda$4959 +
[O~{\sc iii}] $\lambda$5007)/$I$(H$\beta$). c) Relation between the
production efficiency of the ionising photons
$\xi_{\rm ion}$ and O$_{32}$. The canonical value of $\xi_{\rm ion}$ by \citet{R13}
is shown by magenta-shaded region.
Meaning of symbols in b) and c) is the same as in a).
\label{fig1}}
\end{figure*}

The location of 11 selected galaxies in the 
[O~{\sc iii}]$\lambda$5007/H$\beta$ -- [N~{\sc ii}]$\lambda$6584/H$\alpha$ 
diagnostic diagram \citep[Baldwin-Phillips-Terlevich (BPT ) diagram, ][]{BPT81}
in Fig~\ref{fig1}a (red
circles) shows that most of these CSFGs are more 
extreme objects compared to the entire sample of CSFGs 
\citep{I16c}. For the sake of comparison we also show a sample of
SFGs by \citet{Ch14} (blue circles). Their spectra were
obtained with high spectral resolution which was sufficient to derive
velocity dispersion of ionised gas from widths of strong 
H$\beta$ and [O~{\sc iii}] $\lambda$5007 emission lines. The solid line 
derived by \citet{K03} separates SFGs from AGNs. 
All selected CSFGs are located in the SFG region, implying that
their interstellar medium is ionised by hot stars in the star-forming regions. 

In Fig.~\ref{fig1}b we compare the locations in the 
O$_{32}$ -- R$_{23}$ 
(R$_{23}$=\{[O~{\sc ii}]$\lambda$3727 + [O~{\sc iii}]$\lambda$4959 + 
[O~{\sc iii}]$\lambda$5007\}/H$\beta$) diagram of high O$_{32}$ sample
with those of entire CSFGs from SDSS.
It is seen that the selected galaxies with their high O$_{32}$ are located
at the extreme of the distribution for SDSS galaxies, implying 
that they may be good LyC leaking candidates. 
In Fig.~\ref{fig1}b, the CSFGs at $z$ $<$ 0.085 on average are
located somewhat apart from other SFGs because of their lower metallicities and
higher O$_{32}$.

The global characteristics of the CSFGs are derived from their SDSS spectra and
{\sl Galaxy Evolution Explorer} ({\sl GALEX}) photometry, and are shown in
Table \ref{tab1}.
The observed fluxes transformed to luminosities and absolute
magnitudes, adopting luminosity distances \citep[NASA/IPAC Extragalactic
Database (NED),][]{W06} derived with
the cosmological parameters $H_0$=67.1 km s$^{-1}$Mpc$^{-1}$, 
$\Omega_\Lambda$=0.682, $\Omega_m$=0.318 \citep{P14}. 
The H$\beta$ luminosities $L$(H$\beta$) were derived from the 
extinction-corrected H$\beta$ fluxes measured in the SDSS spectra. Additionally,
$L$(H$\beta$) were also corrected for aperture effects using the relation 
2.512$^{r({\rm ap})-r}$, where $r$ and $r$(ap) are respectively the SDSS 
$r$-band total magnitude and the magnitude 
within the round spectroscopic SDSS 3$''$ aperture
for galaxies in the Data Release 9 and 2$''$ aperture for galaxies
in the later SDSS releases. We have also derived extinction-corrected 
absolute SDSS $g$-band and {\sl GALEX} far-UV (FUV) band magnitudes.

The galaxy stellar masses are derived by 
fitting the spectral energy distribution (SED) of the SDSS spectra 
corrected for extinction as derived from the observed hydrogen Balmer decrement
of the same spectra and for the contribution of nebular emission
\citep{I11}. 

\input{tab_obs.tex}

It is seen from Table \ref{tab1} that the selected galaxies are dwarf systems
with low stellar masses $M_\star$ $\la$ 10$^8$ M$_\odot$. 
These extreme CSFGs with low extinction [average $C$(H$\beta$) =
0.084, Table \ref{taba1}] are 
metal-poor, with oxygen abundances 12+log(O/H) between 7.53 and 
7.99 with average value of 7.77 (see also section \ref{sec:abundances}).
Finally, the high equivalent widths of the H$\beta$ emission line in our 
CSFGs (see section \ref{sec:observations}) imply a very high
production efficiency of the ionising photons $\xi_{\rm ion}$ (Fig.~\ref{fig1}c).
These properties are similar to those of the dwarf galaxies at high redshift
thought to be the main contributors to the reionisation of the Universe at
redshifts $z$ $>$ 5.

\section{Observations and data reduction}\label{sec:observations}

We have obtained Xshooter spectrophotometric observations of the galaxies
listed in Table \ref{tab1} in the wavelength range 3200 -- 24000\AA\ with
the three-arm spectrograph splitting emission from the object into three
beams, UVB, VIS, and NIR. The set-ups with the slit of
1\arcsec$\times$11\arcsec\ and resolving power of $R$ = 5400 for the UVB arm,
0\farcs9$\times$11\arcsec\ and $R$ = 8900 for the VIS arm and
0\farcs9$\times$11\arcsec\ and $R$ = 5600 for the NIR arm
were used. The journal of observations is given in Table \ref{tab2}.
Spectra of the spectrophotometric standard stars, obtained during the same
nights, were used for flux calibration and 
correction for telluric absorption lines in the red part of the spectra. 
Additionally, calibration frames of biases, flats and thorium+argon comparison
lamps were obtained during the same nights.

Bias subtraction in the UVB and VIS arms, dark subtraction in the NIR arm,
flat field correction, wavelength and flux calibration, 
and night sky background subtraction were
done with {\sc iraf}. Cosmic ray hits were manually
removed from the background-subtracted frames which were then flux-calibrated. 
Individual subexposures for each object were co-added. Finally,
one-dimensional spectra were extracted in apertures 1\farcs0$\times$2\farcs0
(UVB arm), 0\farcs9$\times$2\farcs0 (VIS arm), and 0\farcs9$\times$2\farcs0
(NIR arm) using the {\sc iraf} {\it apall} routine. All galaxies are
very compact with the full width at half maximum of $\sim$1\arcsec\ in the
$r$-band on SDSS images (Table~\ref{tab2}). Therefore,
apertures for extraction of one-dimensional spectra include almost all
emission from the galaxies.
An example of resulting rest-frame spectra is shown in Fig.~\ref{figb1}.
Strong narrow emission lines are present in the spectrum, suggesting active star
formation.

Emission-line fluxes were measured using the {\sc iraf} {\it splot} routine. 
The errors of the line fluxes were calculated from the
photon statistics in the non-flux-calibrated spectra and adding a 
relative error of 1\% in the 
absolute flux distribution of the spectrophotometric standards.
The line flux errors were propagated in the calculations of the elemental 
abundance errors. 
We note that the strong [O~{\sc iii}]~$\lambda$4363 emission line is present in
all spectra, allowing the reliable electron temerature and abundance
determination. 

\begin{figure*}[t]
\centering
\includegraphics[angle=0,width=0.85\linewidth]{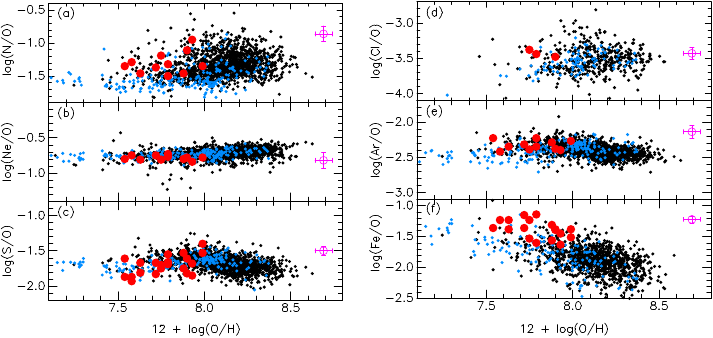}
\caption{Dependencies of the element abundance ratios on the oxygen abundance
which are derived using \citet{I06} prescriptions.
Red symbols are galaxies from this paper. Blue symbols are dwarf
star-forming galaxies observed with various
telescopes for the determination of the helium abundance \citep{I14b} and black
symbols are SDSS star-forming galaxies in which [O~{\sc iii}] $\lambda$4363
emission line is detected at the level above 4$\sigma$. Magenta open circles
with error bars are solar values \citep{L10}.}
\label{fig2}
\end{figure*}

\begin{figure*}[t]
\centering
\includegraphics[angle=0,width=0.80\linewidth]{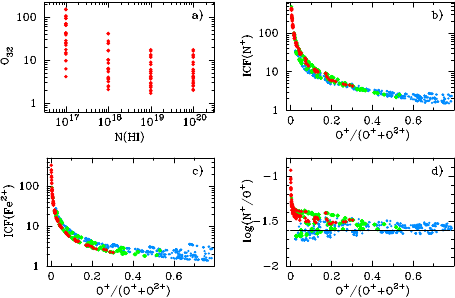}
\caption{a) Dependence of O$_{32}$ ratio on column density of neutral
hydrogen $N$(H~{\sc i}) in the density-bounded H~{\sc ii} region models.
$N$(H~{\sc i}) is expressed in cm$^{-2}$. b) and c) Relations between the
ionisation correction factors $ICF$(N$^+$) and $ICF$(Fe$^{2+}$) and O$^+$
abundance fraction. d) Dependence of N$^+$/O$^+$ abundance ratio on O$^+$
abundance fraction. Black horizontal line indicates the input value of
log(N/O)=--1.6 used in {\sc cloudy} calculations.
Red, blue and green symbols in b) -- d) are density-bounded,
ionisation-bounded models of H~{\sc ii} regions, and H~{\sc ii} regions
with varying covering factor, respectively.}
\label{fig3}
\end{figure*}

The observed fluxes were corrected for extinction with the extinction 
coefficient $C$(H$\beta$),  derived from the observed decrement of the
hydrogen Balmer emission lines.
We have adopted the \citet{C89} reddening law with 
$R(V)$~=~$A(V)$/$E(B-V)$ = 3.1, where $A(V)$ and $E(B-V)$ are respectively the 
total extinction in the $V$ band and the selective extinction.
For $R(V)$ = 3.1, the total extinction is linked to $C$(H$\beta$) by the
relation $A(V)$~=~2.11$\times$$C$(H$\beta$). 
The extinction-corrected emission line fluxes $I$($\lambda$)
relative to the H$\beta$ fluxes multiplied by 100, the extinction coefficients
$C$(H$\beta$), the rest-frame equivalent widths EW(H$\beta$) and the observed
H$\beta$ fluxes $F$(H$\beta$) are listed in Table~\ref{taba1}. We note that the
EW(H$\beta$)s of the studied CSFGs are high, $\sim$~200 -- 390\AA\
(Table \ref{tab1}). Most of them are
among the highest ever measured in spectra of SFGs, indicating
the bursting nature of the star formation and very young burst ages ($<$ 3 Myr).

\section{Element abundances}\label{sec:abundances}

\subsection{Heavy elements}

To determine heavy element abundances, we follow the procedures of \citet{I06}.
We adopt a two-zone photoionised H~{\sc ii} region model: a high-ionisation
zone with the temperature $T_{\rm e}$(O~{\sc iii}), where [O~{\sc iii}],
[Ne~{\sc iii}], and [Ar~{\sc iv}] lines originate, and a low-ionisation zone 
with the temperature $T_{\rm e}$(O~{\sc ii}), where [N~{\sc ii}], 
[O~{\sc ii}], [S~{\sc ii}],
and [Fe~{\sc iii}] lines originate. As for the [S~{\sc iii}] and [Ar~{\sc iii}]
lines, they originate in the intermediate zone between the
high- and low-ionisation regions. The temperature $T_{\rm e}$(O~{\sc iii})
is calculated using the 
[O~{\sc iii}] $\lambda$4363/($\lambda$4959 + $\lambda$5007) ratio.
To calculate the electron temperatures $T_{\rm e}$(O~{\sc ii}) and
$T_{\rm e}$(S~{\sc iii}), we have used the expressions of \citet{I06}.
We note that uncertainties in the determination of $T_{\rm e}$(O~{\sc ii})
do not change the total oxygen abundance because the fraction of O$^+$ ion in
our objects is low.
The electron number densities $N_{\rm e}$(O~{\sc ii}) and $N_{\rm e}$(S~{\sc ii})
are derived from the [O~{\sc ii}] $\lambda$3726/$\lambda$3729
and [S~{\sc ii}] $\lambda$6717/$\lambda$6731 emission line ratios, respectively.
The emission-line errors were propagated for the determination of
electron number density and electron temperature, ionisation correction
factors, ionic and total element abundances, and emission-line velocity
dispersions. However, they do not take into account systematic effects
introduced e.g. by uncertainties in the determination of ionisation correction
factors.

The electron temperatures $T_{\rm e}$(O {\sc iii}) in the studied CSFGs are high
(Table \ref{taba2}), ranging from $\sim$15000K to $\sim$20000K. 
The electron number densities $N_{\rm e}$(O {\sc ii}) and $N_{\rm e}$(S {\sc ii}),
characteristic of the low-ionisation zone in the H {\sc ii} region, are in
general agreement within the errors and for some galaxies are relatively high.

Ionic and total heavy element abundances are derived using \citet{I06} relations
and are
presented in Table~\ref{taba2} together with the ionisation correction factors
ICF for unseen stages of ionisation. The abundance ratios N/O, Ne/O, S/O, Cl/O,
Ar/O, and Fe/O
are shown in Fig.~\ref{fig2}. Except for nitrogen and iron
(Fig.~\ref{fig2}a and \ref{fig2}f), the abundance ratios 
in the CSFGs studied in this paper (red filled circles) are similar to
those derived in a sample of BCDs with high signal-to-noise ratio spectra
\citep{I06,I14b} shown by blue filled circles and to SDSS galaxies (black filled
circles) in which [O~{\sc iii}] $\lambda$4363 emission line is detected at
the level above 4$\sigma$. The nitrogen and iron to oxygen abundance ratios in
our sample of CSFGs are higher than in comparison sample of galaxies by
\citet{I06} with milder ionisation radiation indicated by lower O$_{32}$ ratios
($<$10) and with somewhat older ages of starbursts indicated by lower
EW(H$\beta$).

Several mechanisms can be proposed to explain the enhanced N/O and Fe/O
abundance ratios in our galaxies with extreme characteristics.
Nitrogen-enriched clumps in inhomogeneous H~{\sc ii} regions with
Wolf-Rayet stellar population likely need to 
be invoked to explain such high N/O abundance ratios at low metallicity
(12+log(O/H) $\sim$ 8) \citep{I06}.
\citet{Ch23}, \citet{Is23} and \citet{M24} proposed to explain high N/O
abundance ratios by evolution of Wolf–Rayet stars or supermassive
(10$^3$ -- 10$^5$ M$_\odot$) stars. \citet{Is22} and \citet{Wa24}
proposed to explain high N/O and Fe/O abundance ratios by
nucleosynthesis from massive hypernovae and pair-instability supernovae
whereas \citet{Ko21} proposed to explain nearly solar Fe/O abundance ratio
in extremely metal-poor galaxies by nucleosynthesis in very massive stars
with masses above 300 M$_\odot$.

\input{tab_cloudy.tex}

However, we point out that both the nitrogen and iron abundances are derived,
respectively, from low-ionisation [N~{\sc ii}]$\lambda$6584, and
[Fe~{\sc iii}]$\lambda$4658, 4988 emission lines. The fraction of N$^+$ and
Fe$^{2+}$ is low in our galaxies with extremely high O$_{32}$ ratios. Therefore,
high ionisation correction factors ICF(N) and ICF(Fe) are needed to convert
ion abundances to total element abundances (Table~\ref{taba2}). These
ionisation correction factors are determined from the ionisation-bounded
models. 
We point out that high O$_{32}$ ratios in our galaxies imply that
the H~{\sc ii} regions in these galaxies might be clumpy or
density-bounded. We verify here how N/O and Fe/O abundance ratios can differ
in clumpy or density-bounded
H~{\sc ii} regions. Clumpiness of the ionised gas is determined by the
filling factor, whereas clumpiness of the neutral gas is defined by the
covering factor ranging from 0 to 1.

We use the {\sc cloudy} v25.00 code \citep{G25}
to calculate a set of uniform spherically-symmetric photoionised H~{\sc ii}
region models with finite column density of neutral hydrogen ranging from
10$^{17}$--10$^{20}$ cm$^{-2}$ or covering factor ranging from 0 (open
geometry) to 1 (closed geometry) and to
compare them with ionisation-bounded H~{\sc ii} region models. We also
adopted the input oxygen abundance 12+log(O/H) = 7.6, which is typical for
our galaxies, and starburst ages of 2 and 3 Myr. For production rate of ionising
photons we adopted two values, $Q_{\rm ion}$ = 10$^{52}$ -- 10$^{53}$ s$^{-1}$
corresponding to the recombination H$\beta$ luminosity of
$\sim$ 10$^{40}$ -- 10$^{41}$ erg s$^{-1}$, which is also typical for our
galaxies. Two other parameters, which we vary in a large range, are electron
number density $N_{\rm e}$ and filling factor $f$. The entire range of input
parameters is  shown in Table \ref{tab3}. We note that the product
$N_{\rm e}f$ determines the compactness of the H~{\sc ii} region model.

Variations of O$_{32}$ ratio on neutral hydrogen column density in our
models is shown in Fig.~\ref{fig3}a. Several values of O$_{32}$ at fixed
$N$(H~{\sc i}) correspond to different values of $N_{\rm e}f$. It is seen that
O$_{32}$ decreases with increasing $N$(H~{\sc i}) at $N$(H~{\sc i}) in
the range of 10$^{17}$ -- 10$^{19}$ cm$^{-2}$ but attains constant values at
higher $N$(H~{\sc i}). The model values of O$_{32}$ at
$N$(H~{\sc i}) = 10$^{17}$ cm$^{-2}$ are too high compared to observed values.
Therefore, translucent models with $N$(H~{\sc i}) $>$ 10$^{17}$ cm$^{-2}$
are more likely.

In Figs.~\ref{fig3}b and \ref{fig3}c we show ionisation correction factors
for nitrogen and iron for both the density-bounded,
clumpy, and ionisation-bounded models of H~{\sc ii} regions. The small
difference between three sets of models suggests that applying $ICF$s
for density-bounded models and relations from \citet{I06} for N$^+$/H$^+$ and
Fe$^{2+}$/H$^+$ abundance ratios would result in small variations of N/H and
Fe/H abundance ratios, not exceeding 0.10 and 0.15 dex, respectively, at
O$^+$/(O$^+$+O$^{2+}$) $>$ 0.05 compared to ionisation-bounded
H~{\sc ii} regions.

We also note that the relation N/O $\approx$ N$^+$/O$^+$ is used in some
studies \citep[e.g. ][]{B20,AC25} since ionisation potential of N is
slightly higher than that of O.
Application of this relation results in $\sim$ 0.05 -- 0.10
higher values of log N/O than those in Table~\ref{taba2}. An exception
is higher N$^+$/O$^+$ ratios in density-bounded models with very high
O$_{32}$ values because of higher cuts of O$^+$ zone compared to that of N$^+$
zone.

Thus, the variations of ICFs as well as of log(N/O) and log(Fe/O) abundance
ratios lead us to conclusion that the discussions about
the possible origin of N and Fe in galaxies with extreme O$_{32}$ should be
considered with caution if the N and Fe abundances are derived using
low-ionisation [N~{\sc ii}]~$\lambda$6584, [Fe~{\sc iii}]~$\lambda$4658,
$\lambda$4988 emission lines. Furthermore, all above abundance determinations
are based on the assumption of ionisation-bounded H~{\sc ii} region models,
whereas high O$_{32}$ values may indicate that H~{\sc ii} regions can be
density-bounded or clumpy, although the ionisation correction factors are
similar in all models if O$^+$/(O$^+$+O$^{2+}$) $>$ 0.05.


\subsection{Helium}

\input{tab_minim.tex}

\input{tabhe2.tex}

The determination of the He abundance in low-metallicity SFGs
requires special detailed consideration. This is because the high
precision of He abundances is needed to derive the primordial He abundance
from these data. Until now almost all He abundance determinations were made
using observations of low-redshift star-forming galaxies
\citep[e.g. ][]{I14b,Ma22,Ya25}. However, with the advent of
{\sl James Webb Space Telescope} ({\sl JWST}),
the He abundance determinations for high-redshift galaxies
with 1.6 $\la$ $z$ $\la$ 3.3 become possible \citep{Be26}.

In this study, using Xshooter
observations, we self-consistently derive the He abundance in
the way used e.g. by \citet{ITL97,IT98,IT04,I14b}. The method is based on
minimisation of differences
in the He mass fractions Y$_{\rm i}$ derived from different He~{\sc i} emission
lines after correction of their intensities for collisional and fluorescent
enhancement, using approximations for both effects on optical depth
$\tau$($\lambda$3889) and the electron number density $N_{\rm e}$(He) in the zone
of He~{\sc i} emission according to \citet{B99}, \citet{B02}, \citet{P12,P13},
and \citet{I13}. Alternative schemes of minimisation were developed
e.g by \citet{Av11,Av12,Av15}, \citet{Ya25}, \citet{Be26}.
We note that the electron number density $N_{\rm e}$(S~{\sc ii})
is used in many papers to correct for collisional enhancement of
He~{\sc i} emission lines. However, emission of
[S~{\sc ii}] $\lambda$6717, 6731 lines is produced in outer region
of the H~{\sc ii} region and in the surrounding H~{\sc i} region. Therefore,
$N_{\rm e}$(S~{\sc ii}) may not be appropriate for the He abundance
determination.

Apart from collisional and fluorescent enhancement of the He~{\sc i} emission
lines we take into account some other
effects deviating hydrogen and helium emission lines from recombination
values including the presence of underlying stellar absorbtion lines,
interstellar extinction and collisional excitation of hydrogen lines.
Finally, we added the mass fraction of He in the He$^{2+}$ stage, using the
intensity of the He~{\sc ii} $\lambda$4686 emission line.

We consider two cases of weighted mean Y determination using two
sets of He~{\sc i} emission lines, one set is with six lines,
He~{\sc i} $\lambda$3889,
4471, 5876, 6678, 7065 and 10830\AA, and the second set is with five lines
excluding He~{\sc i} $\lambda$10830\AA\ emission line. It is important to
use the He~{\sc i} $\lambda$10830\AA\ emission line because this line
strongly depends on the electron number density.

\begin{figure}[t]
\centering
\includegraphics[angle=0,width=0.95\linewidth]{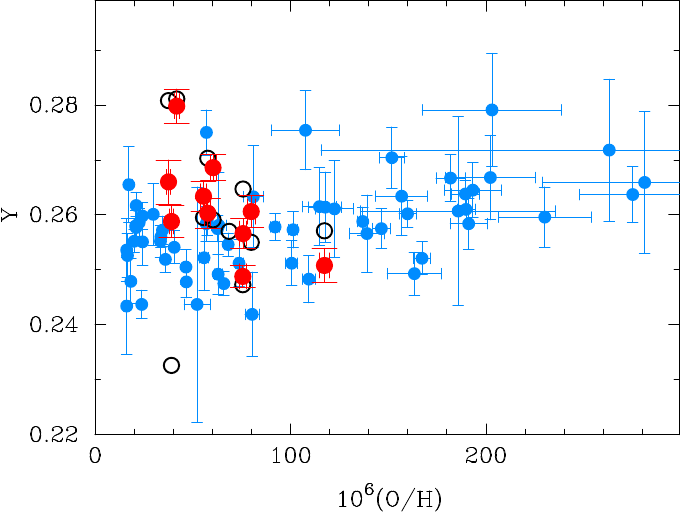}
\caption{Dependence of the helium mass fraction Y on the oxygen abundance
O/H for star-forming galaxies. The galaxies from \citet{I14b} with Ys
derived using six He~{\sc i} emission lines are shown by blue
symbols and galaxies from this paper are shown by red symbols for galaxies
with Ys derived using six He~{\sc i} emission lines, and by black symbols for
galaxies with Ys derived using five He~{\sc i} emission lines.}
\label{fig4}
\end{figure}

The range of parameters $T_{\rm e}$(He), $N_{\rm e}$(He),
$\tau$($\lambda$3889) used in the He abundance minimisation is shown
in Table~\ref{tab4} and their values for the minimum deviation of Ys for
different He~{\sc i} lines from the weighted mean Y(mean) are shown in
Table~\ref{taba2}. The values of Y for different He~{\sc i} emission lines
are presented in Table~\ref{tab5}. They vary in a large range from line
to line and from galaxy to galaxy mainly because of large statistical
uncertainties, but weighted mean Y(mean) vary in much narrower range and
have small statistical errors. These values are similar to the values of
Y(mean) derived e.g. by \citet{I14b} for a large sample of galaxies with
similar physical characteristics using the same minimisation technique
(Fig.~\ref{fig4}). We also note that the dispersion of Y values derived
using six He~{\sc i} emission lines is somewhat lower than that derived using
five He~{\sc i} emission lines (red and black symbols in Fig.~\ref{fig4},
respectively). Earlier this fact was noted by \citet{I14b} who used much larger
sample of galaxies. We note that the intensity of the
He~{\sc i} $\lambda$10830 emission line is affected by strong telluric
absorption in spectra of two galaxies, J0009+0234 and J1310+0852.
Therefore, its intensity relative to H$\beta$ in spectra of these galaxies
is unphysically low, below the recombination value of $\sim$ 0.2
(Table~\ref{taba1}).
Consequently, the Y(10830)s in these galaxies are considerably lower than
for other galaxies. We also note that the intensity of this line might be
reduced by dust absorption because of non-zero optical depth and multiple 
scattering. We can not estimate this reduction
but it is likely not high because of the low metallicity in our galaxies.

\section{Kinematics of the ionised gas}\label{sec:kinematics}

The large wavelength coverage, medium spectral resolution of the
Xshooter spectrograph,
and relatively high apparent brightness of our extreme galaxies allow us to
derive velocity dispersions $\sigma$ for a large number of emission lines from
ions with various ionisation potentials originating in different parts of
H~{\sc ii} regions, and to study kinematics of the ionised gas. Previously, 
kinematics of ionised gas in a large sample of low-redshift high-excitation
H~{\sc ii} regions has been studied e.g. by \citet{Ch14} with various
telescopes using spectroscopic observations with resolving power $R$ $>$ 10000.
They aimed to derive the relation between
the extinction-corrected H$\beta$ luminosity $L$(H$\beta$) and velocity
dispersion $\sigma$(H$\beta$) to use this relation for determination of the
Hubble constant $H_0$. However, \citet{Ch14} considered only two strong lines,
H$\beta$ and [O~{\sc iii}] $\lambda$5007\AA.

\begin{figure}[t]
\centering
\includegraphics[angle=0,width=0.99\linewidth]{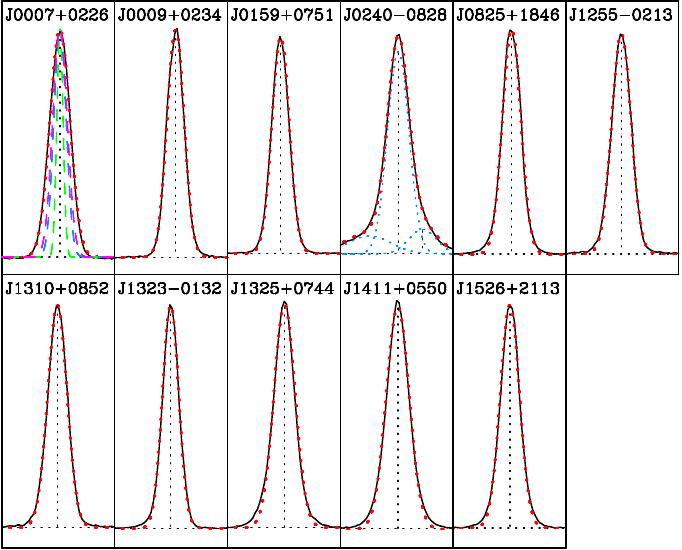}
\caption{Profiles of the H$\beta$ emission line in our galaxies. The observed
profiles are shown by black solid lines. All profiles can be fitted by single
Gaussians (red dotted lines) except for J0240$-$0828. The profile
of the latter galaxy is fitted by three Gaussians (blue dotted lines) with
the total fitted profile shown by the red dotted line. For the galaxy
J0007$+$0226 by blue, green and magenta dashed lines are shown instrumental,
thermal and macroscopic (turbulent) profiles, respectively.}
\label{fig5}
\vspace{0.2cm}
\centering
\includegraphics[angle=0,width=0.99\linewidth]{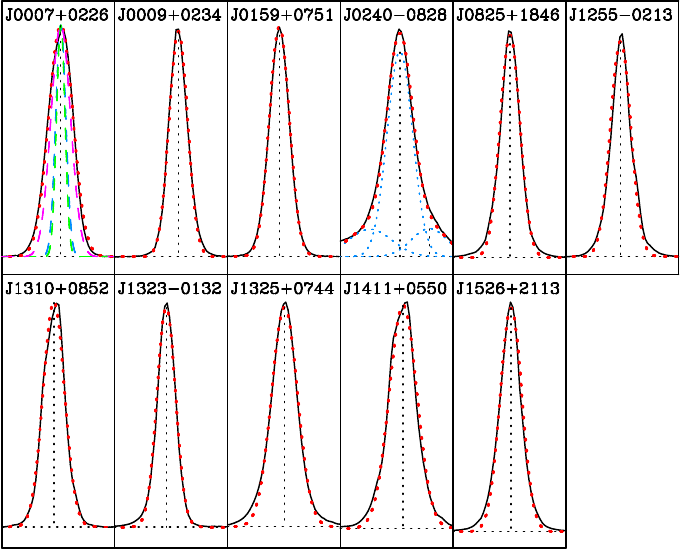}
\caption{Same as in Fig.~\ref{fig5} but for the H$\alpha$ profile.}
\label{fig6}
\end{figure}
\begin{figure}[t]
\centering
\includegraphics[angle=0,width=0.95\linewidth]{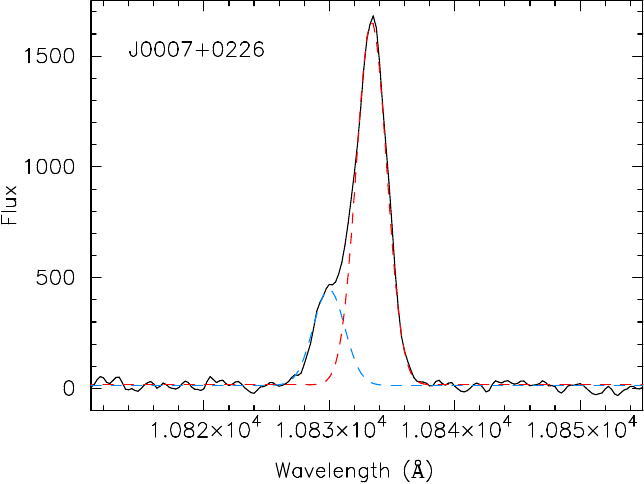}
\caption{Two Gaussian fitting (dashed lines) of the
He~{\sc i} $\lambda$10830\AA\ emission
line profile in the spectrum of J0007+0226 (black solid line).}
\label{fig7}
\end{figure}

\input{tab_broad1_6.tex}

All our galaxies are very compact and are unresolved in their SDSS images.
Furthermore, some objects were imaged with the {\sl HST}, again showing
very compact structure. This implies that their emission lines
originate mainly in a single H~{\sc ii} region and thus can be fitted
by single Gaussians. However, very weak broad emission of strongest
lines is also present in some of our galaxies, most commonly of the
[O~{\sc iii}] $\lambda$5007\AA\ line. In these cases we applied
two Gaussian fitting and compared widths of narrow components obtained by this
method and one Gaussian fitting. We find that widths of narrow components
derived by these two methods differ only by less than
0.1\%. This is because the fluxes of narrow components are by two orders
of magnitude higher than those of broad components. Therefore, one Gaussian
fitting is applicable to our galaxies and we adopt it in this paper, excluding
He~{\sc i} $\lambda$10830\AA, for which two Gaussian fitting is adopted
because of broading due to radiative transfer (see below). We note that
the broad components are not central in this paper and their analysis is not
present here.

A fitting with a single Gaussian of strong emission lines
H$\beta$ and H$\alpha$ in all studied galaxies is shown
in Figs.~\ref{fig5} and \ref{fig6}. It is clear from these Figures
that emission in the line is very well fitted with a single Gaussian,
only very slightly deviating from the observed line profile in the weak wings.
An exception is the galaxy J0240$-$0828, for which the fit by three
Gaussians is adopted with one dominant Gaussian. The contribution of other
two Gaussians to the line emission is small. Therefore, for
clarity, we adopt the simplest model with a single Gaussian for ten galaxies
and a dominant Gaussian for J0240$-$0828.

We derive the observed velocity dispersion $\sigma_{obs}$ from the relation
$\sigma_{obs}$~=~FWHM/2.35 
assuming that the line profile is a Gaussian.
FWHM is the rest-frame full width of the line at half maximum.

The velocity dispersion $\sigma_{obs}$ has been corrected for thermal
($\sigma_{th}$) and instrumental ($\sigma_{in}$) broadening to derive the final
dispersion $\sigma$, characterising macroscopic turbulent motion, in accord with
\citet{Ch12}, using relation
\begin{equation}
\sigma=(\sigma_{obs}^2-\sigma_{th}^2-\sigma_{in}^2)^{1/2}. \label{sigma}
\end{equation}
The instrumental dispersion in each arm is derived from the resolving
power according to
\begin{equation}
\sigma_{in} = \frac{c}{2.35R},
\end{equation}
where $R$ is the resolving power and $c$ is a speed of the light.
Adopting $R$ = 5400, 8900 and 5600 for UVB, VIS and NIR arms, we obtain
the respective $\sigma_{in}$ = 23.6 km s$^{-1}$, 14.3 km s$^{-1}$, and
22.2 km s$^{-1}$. We confirmed these values of $\sigma_{in}$ from the
measurements of emission line widths of Thorium-Argon comparison spectra
in all orders of the UVB and VIS arms, and of night-sky spectra in
all orders of the NIR arm, obtained
with the same slit widths as those used for obtaining spectra of the galaxies.
Furthermore, very similar velocity dispersions $\sigma$ of hydrogen
Balmer and Paschen emission lines observed in all UVB, VIS and NIR arms
(see later) implies that dispersions of instrumental profiles in different arms
were derived correctly.

\begin{figure*}[t]
\centering
\includegraphics[angle=0,width=0.80\linewidth]{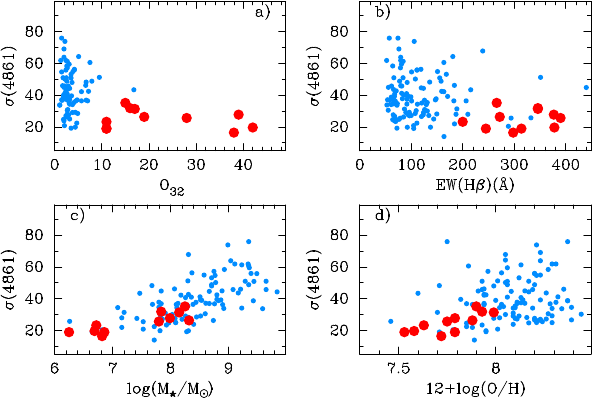}
\caption{Relations between the velocity dispersion of the H$\beta$ emission
line (in km s$^{-1}$) and (a)
O$_{32}$ = [O~{\sc iii}]5007/([O~{\sc ii}]3726 + [O~{\sc ii}]3729),
(b)~equivalent width EW(H$\beta$) of the H$\beta$ emission line,
(c) log($M_\star$/M$_\odot$), and (d) oxygen abundance 12+log(O/H).
Our galaxies are represented by
red filled circles and blue filled circles represent galaxies by \citet{Ch14}.}
\label{fig8}
\end{figure*}

The thermal dispersion is derived from the relation
\begin{equation}
\sigma_{th}=\sqrt{\frac{kT_{\rm e}}{Am_p}},
\end{equation}
assuming a Maxwellian velocity distribution, where $k$ is the Boltzman constant,
$m_p$ is the proton mass, $A$ is the
atomic number of the ion, and $T_{\rm e}$ is the electron temperature derived
from the spectrum (Table \ref{taba2}). We adopt $T_{\rm e}$(O~{\sc iii}) for
all emission lines considered below except for [O~{\sc ii}] $\lambda$3726\AA,
$\lambda$3729\AA, [S~{\sc ii}] $\lambda$6717\AA,
$\lambda$6731\AA, for which we use $T_{\rm e}$(O~{\sc ii}). The thermal
dispersion is comparable to instrumental dispersion for the VIS arm but
it is considerably smaller for the UVB and NIR arms.

There are two He~{\sc i} transitions from the metastable 2$^3$S state, namely
He~{\sc i} $\lambda$3889\AA\ and $\lambda$10830\AA, which may be subject to
resonant scattering because of non-zero optical depth. Usually, the
He~{\sc i} $\lambda$3889\AA\ emission line is considered in He abundance
determination as a measure of the optical depth \citep[e.g. ][]{B99,B02}.
Recently \citet{Dr26} considered the radiative transfer in the
He~{\sc i} $\lambda$10830\AA\ emission line.
He predicts the complex He~{\sc i} 10830\AA\ profile for typical
conditions in the H~{\sc ii} region consisting of two strong peaks separated
by $\sim$ 75 -- 140 km s$^{-1}$ and some other weaker features.
This mechanism makes the He~{\sc i} $\lambda$10830 emission line in the Xshooter
spectra somewhat broadened and asymmetric.
Therefore, for this line we applied model with two Gaussian fitting to
derive separation between the peaks. Results of this modelling are present in
Table \ref{tab7}. An example of the fitting for the galaxy J0007+0226 is shown
in Fig.~\ref{fig7}. The two Gaussian fitting was not possible for four
galaxies. In two galaxies, J0009+0234 and J1310+0852, the line is subject
to telluric absorption and the profile is symmetric in other two galaxies.
It is seen from the Table that separation between the peaks in spectra of other
galaxies is consistent with the peak separation predicted by \citet{Dr26} for
resonant scattering. For further comparison of the
He~{\sc i} $\lambda$10830\AA\ emission line with other emission lines we adopt
Gaussian characteristics only for its brightest component.

\input{tab10830.tex}

Details of H$\beta$ and H$\alpha$ profile fitting for the galaxy J0007$+$0226
are shown in Figs.~\ref{fig5} and \ref{fig6}, where the observed profiles and
their one-Gaussian fits are represented by black solid and red dotted
lines, respectively. By blue, green and magenta dashed lines are shown
instrumental, microscopic (thermal) and macroscopic (turbulent) components,
respectively. The observed wavelength of the H$\beta$ emission line in spectra
of all our galaxies falls to the wavelength range of the UVB arm, whereas the
H$\alpha$ emission line was observed
at wavelengths covered by the VIS arm. Therefore, the H$\beta$ profile is only
partly resolved because of the relatively broad instrumental profile
(Fig.~\ref{fig5}). The resolution is much better for the H$\alpha$ profile
because of almost twice narrower instrumental profile (Fig.~\ref{fig6}).

Velocity dispersions $\sigma$ of the 17 brightest lines in the spectra of all
11 selected galaxies are shown in Table \ref{tab6}. Some bright emission lines
are not considered, e.g. P$\alpha$ $\lambda$18756\AA,
[S~{\sc iii}] $\lambda$9069\AA\ and $\lambda$9531\AA, because
they are observed at the wavelengths with relatively strong telluric absorption.

In Fig.~\ref{fig8} we compare velocity dispersions $\sigma$(H$\beta$) and
various other characteristics of our extreme galaxies observed with the
Xshooter and galaxies by \citet{Ch14}. It is seen in Fig.~\ref{fig8}a that
the sample by \citet{Ch14} is characterized by low O$_{32}$ ratios, not
exceeding the value of 4 for most galaxies. Contrary to that galaxies observed
with Xshooter are much more extreme, with O$_{32}$ ratio extending to very high
values from 10 to more than 40. Equivalent widths of H$\beta$ emission line in most galaxies
from the sample by \citet{Ch14} are also much lower than those in galaxies
from our sample (Fig.~\ref{fig8}b). EW(H$\beta$) is a characteristic of the
H~{\sc ii} region age. Therefore, lower EW(H$\beta$)s indicate older age of
H~{\sc ii} regions in the vast majority of galaxies by \citet{Ch14}.
Besides that, the contribution
of underlying older stellar population to emission in the H~{\sc ii} region
of most galaxies from the \citet{Ch14} sample is higher. At fixed $\sigma$(4861)
our galaxies are systematically less massive (Fig.~\ref{fig8}c) and more
metal-poor (Fig.~\ref{fig8}d). It is worth to emphasize that all our galaxies
have low velocity dispersion of the H$\beta$ emission line, and for such low
dispersion we increase the statistics in the range of low metallicity,
low stellar mass and high H$\beta$ equivalent width.
We find that $\sigma$(H$\beta$) increases with the stellar mass and oxygen
abundance and slightly decreases with O$_{32}$. On the other hand, there is no
correlation between $\sigma$(H$\beta$) and EW(H$\beta$).

\begin{figure*}[t]
\centering
\includegraphics[angle=0,width=0.95\linewidth]{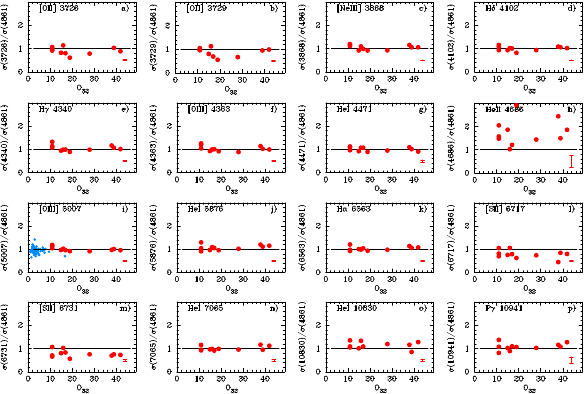}
\caption{The ratios of velocity dispersions of various lines to the velocity
dispersion of the H$\beta$ emission line in function of O$_{32}$ ratios
(red filled circles) with typical errors shown in the right lower corners.
Blue filled circles in panel (i) are the galaxies by \citet{Ch14}.}
\label{fig9}
\end{figure*}

The Xshooter observations allow us to derive velocity dispersions not only of
H$\beta$, H$\alpha$, and [O~{\sc iii}]$\lambda$5007\AA\ emission lines, but
also of several other strong recombination and collisionally-excited emission
lines of ions with different ionisation potentials. This makes possible an
investigation of kinematic properties in various parts of H~{\sc ii} regions.
In the past, similar studies have been done for some blue compact dwarf and
green pea galaxies \citep[e.g. ][]{J09,A12,F11,H12}. However, no such study has
been done before for CSFGs with extremely high O$_{32}$ ratios.
For convenience, to compare velocity dispersion $\sigma$($\lambda$) of
different emission lines, we will use the ratio of velocity dispersions of a
particular line to that of the H$\beta$ emission line,
$\sigma$($\lambda$)/$\sigma$(4861).
The dependence of $\sigma$($\lambda$)/$\sigma$(4861) on O$_{32}$ ratios,
stellar masses, equivalent widths of the H$\beta$ emission line and oxygen
abundances are shown in Figs.~\ref{fig9}, \ref{figb2}, \ref{figb3} and
\ref{figb4}, respectively. We also show data for galaxies from \citet{Ch14} by
blue filled circles in
panel (i) of these Figures. We note that O$_{32}$ is a proxy of the ionisation
parameter and it varies in a wide range. The $\sigma$($\lambda$)/$\sigma$(4861)
ratios are nearly constant and equal to $\sim$ 1 for
[Ne~{\sc iii}] $\lambda$3868\AA, [O~{\sc iii}] $\lambda$4363\AA, $\lambda$5007\AA\ emission lines, most of
He~{\sc i} and all hydrogen lines. This indicates that all mentioned
lines originate in the same parts of H~{\sc ii} regions. On the other hand,
the ratios $\sigma$($\lambda$)/$\sigma$(4861) considerably deviate from unity
for [O~{\sc ii}]$\lambda$3726\AA, $\lambda$3729\AA, [S~{\sc ii}]$\lambda$6717\AA, $\lambda$6731\AA,
He~{\sc ii} $\lambda$4686\AA\ and He~{\sc i} $\lambda$10830\AA\ emission lines.

The most deviant is the He~{\sc ii} $\lambda$4686\AA\ emission line with velocity
dispersion which is on average $\sim$ 1.5 times larger than $\sigma$(4861) of
the H$\beta$ emission line (Figs.~\ref{fig9}h, \ref{figb2}h, \ref{figb3}h and
\ref{figb4}h).
The He~{\sc ii} $\lambda$4686\AA\ emission line is likely
produced in the inner part of the H~{\sc ii} region and can be broadened by
dynamical processes such as stellar winds of massive stars, massive X-ray
binaries and/or shocks in expanding supernova remnants from current and/or past bursts of
star formation. However, the presence of AGN can likely be excluded, because
the high-ionisation emission line [Ne~{\sc v}] $\lambda$3426\AA\
is detected in spectra of only two galaxies, J0009+0234 and J0240$-$0828, with
the $I$([Ne~{\sc v}] $\lambda$3426\AA)/$I$(He~{\sc ii} $\lambda$4686\AA) ratios
of $\sim$ 0.4 and $\sim$ 0.1. These low ratios are typical for star-forming
galaxies with the detected [Ne~{\sc v}] emission, contrary to high ratios of
$\sim$ 1--3 in Seyfert 2 galaxies \citep[e.g. ][]{I12,I21c,TI05}.

Similar appearance of the He~{\sc ii} $\lambda$4686\AA\ emission line has been
found e.g. by \citet{I06b} in the blue compact dwarf galaxy SBS 0335$-$052E.
They have found that the width of this line in the galaxy is 1.5 -- 2 times
greater than the width of other lines. Furthermore, \citet{TI05} detected
weak [Ne~{\sc v}] $\lambda$3426\AA\ emission line in SBS 0335--052E with
intensity, typical for star-forming galaxies.
However, we note that, at present, the origin
of high-ionisation He~{\sc ii} and [Ne~{\sc v}] emission lines
in SBS\,0335$-$052E (and other star-forming galaxies with detected [Ne~{\sc v}]
emission) remains unsolved, despite the considerable efforts \citep[e.g. ][]{Mi25}.

On the other hand, the $\sigma$([O~{\sc ii}] 3726)/$\sigma$(H$\beta$ 4861),
$\sigma$([O~{\sc ii}] 3729)/$\sigma$(H$\beta$ 4861),
$\sigma$([S~{\sc ii}] 6717)/$\sigma$(H$\beta$ 4861) and
$\sigma$([S~{\sc ii}]~6731)/$\sigma$(H$\beta$ 4861) ratios 
are systematically lower than unity and likely do not depend on O$_{32}$
(Figs.~\ref{fig9}a, \ref{fig9}b, \ref{fig9}l, \ref{fig9}m)
and EW(H$\beta$) (Figs.~\ref{figb3}a, \ref{figb3}b, \ref{figb3}l, \ref{figb3}m).
This appearance can be explained by the
fact that low-ionisation ions O$^+$ and S$^+$ exist in the outer part of the
H~{\sc ii} region which is likely less disturbed by dynamical processes caused
by massive stars in the central cluster, resulting in the relatively low
velocity dispersion. Similarly, the
$\sigma$([O~{\sc ii}]~3726)/$\sigma$(H$\beta$~4861),
$\sigma$([O~{\sc ii}]~3729)/$\sigma$(H$\beta$~4861),
$\sigma$([S~{\sc ii}]~6717)/$\sigma$(H$\beta$~4861) and
$\sigma$([S~{\sc ii}]~6731)/$\sigma$(H$\beta$~4861) ratios do not depend
on the stellar mass (Figs.~\ref{figb2}a, \ref{figb2}b, \ref{figb2}l, \ref{figb2}m)
and oxygen
abundance (Figs.~\ref{figb4}a, \ref{figb4}b, \ref{figb4}l, \ref{figb4}m).

We note that the [N~{\sc ii}] $\lambda$6584 emission line in our galaxies is
generally weaker than [S~{\sc ii}] emission lines (Table~\ref{taba1}). We
derived velocity dispersions $\sigma$ for this line only in six galaxies with
highest signal-to-noise ratios in the continuum and with the
[N~{\sc ii}]~$\lambda$6584 profiles that can be fit by a single Gaussian.
We find that $\sigma$([N~{\sc ii}]~$\lambda$6584)/$\sigma$(H$\beta$) in these
galaxies are systematically lower than unity, similarly to that for
[O~{\sc ii}] and [S~{\sc ii}] emission lines, meaning that they are emitted
in nearly same outer zones of H~{\sc ii} region.

As for infrared hydrogen line P$\gamma$ 10941\AA, its measured velocity
dispersion may be affected by
telluric absorption. But this line in spectra of our galaxies is
located at wavelengths with relatively low
telluric absorption resulting in $\sigma$(10941)/$\sigma$(4861) $\sim$ 1
(Fig.~\ref{fig9}n). 

Interesting features are found for He~{\sc i} emission lines. We present data
only for emission lines arising in transitions between triplet states of
ortho-He (He~{\sc i} 4471, 5876, 7065, 10830 (Table~\ref{tab6},
Figs.~\ref{fig9}, \ref{figb2}, \ref{figb3}, \ref{figb4}). Emission lines arising
in the transitions between singlet
states (para-He) not blended with other lines are also present in spectra of
our galaxies (e.g. He~{\sc i} 4922, 5016, 6678) but they are much weaker than
the triplet lines. Therefore, their velocity dispersions are subject to higher
uncertainties and we do not consider them.

Some of the He~{\sc i} emission lines ($\lambda$7065 and $\lambda$10830) can be
subject to telluric absorption.
Regarding to the He~{\sc i} $\lambda$10830 emission
line this is true for two objects, J0009+0234 and J1310+0852. The intensities
of the He~{\sc i} $\lambda$10830 emission line in these two galaxies
(Table~\ref{taba1}) are lower than the predicted recombination value
($I$(He~{\sc i} 10830)/$I$(H$\beta$) $\sim$ 0.2), whereas the intensities
of nearby hydrogen P$\gamma$ $\lambda$10941 emission line are consistent
with the recombination value ($I$(P$\gamma$ 10941)/$I$(H$\beta$) $\sim$ 0.07).
But in other galaxies the He~{\sc i} emission lines $\lambda$7065 and
$\lambda$10830 are located at wavelengths free of telluric
absorption. We also checked whether possible systematic
uncertainties in the wavelength calibration may influence the velocity
dispersions. For this we compare widths of particular emission line, not
only of He~{\sc i} but of ions of other elements, measured in two
adjacent orders of echelle spectrum, where it is possible. We find
that differences in widths of the same line in the two orders
are in the range between 2\% and 5\%.

It is seen in Figs.~\ref{fig9}g and \ref{figb2}g that velocity dispersions of
He\,{\sc i}\,$\lambda$4471, $\lambda$5876, $\lambda$7065 are almost equal
to that of the H$\beta$ emission line indicating that collisional and
fluorescent enhancement have minor effect on broadening of these lines.
On the other hand, the
He\,{\sc i}\,$\lambda$10830 emission line is on average by $\sim$20\%
broader than H$\beta$ emission line. The higher $\sigma$(10830) is real and is
not due to the uncertainties of the data reduction because the velocity
dispersion of nearby P$\gamma$ $\lambda$10941 emission line is the same as that
for H$\beta$ (Fig.~\ref{fig9}o -- \ref{figb2}o). In part, this broadening is
caused by radiative scattering in the He\,{\sc i}\,$\lambda$10830 emission
line. Another emission line which is subject to radiative scattering is
the He~{\sc i} $\lambda$3889 emission line. However, this line is blended with
the H~{\sc i} $\lambda$3889 emission line. The line separation in the blend is
0.4\AA, corresponding to velocity separation of $\sim$30 km s$^{-1}$, and the
blend is characterised by slightly asymmetric profile. We
measured the velocity dispersion of the blend adopting one Gaussian
fitting and found that it is above 30 km s$^{-1}$ in all galaxies. Thus,
broadening of the blend is likely caused by wavelength difference of the
H~{\sc i} and He~{\sc i} lines. Perhaps, additionally, the
He\,{\sc i}\,$\lambda$10830 emission line tends to be emitted
in somewhat different, likely inner regions with somewhat higher density
compared to other He~{\sc i} emission lines.

Apparently, inner regions are characterised by higher turbulent velocities.
This difference between
He\,{\sc i}\,$\lambda$10830 and other He~{\sc i} lines, though small, may
potentially pose the problem in the determination of the He abundance weighted
by several He\,{\sc i} emission lines because they are emitted in somewhat
different parts of the H~{\sc ii} region. The best lines for
the He abundance determination might be He~{\sc i} $\lambda$4471 and likely
He~{\sc i} $\lambda$6678 because they are least dependent on collisional and
fluorescent enhancement and their velocity dispersions are equal to those of
H$\beta$, at least for He~{\sc i}~$\lambda$4471 line. However, these lines are
weak ($\sim$3--4\% of the H$\beta$ intensity), therefore spectra with high
signal-to-noise ratio are needed to derive their intensities with good
precision. Furthermore, the intensity of the He~{\sc i}~$\lambda$4471\AA\
emission line can be affected by underlying stellar He~{\sc i} absorption line
in larger extent than other He~{\sc i} emission lines.

\section{Conclusions}\label{sec:conclusions}

In this paper we present VLT/Xshooter 
spectrophotometric observations of 11 low-mass compact star-forming
galaxies (CSFGs) at $z$ $<$ 0.085 with extremely high O$_{32}$ = 11 -- 42
(`high O$_{32}$ sample').
Our goal here is to study the physical conditions and chemical composition and
to investigate kinematic properties of galaxies in various hydrogen, helium,
oxygen, neon and sulphur emission lines. 
Our main results are as follows.

1. All spectra show strong emission lines implying the presence of a very 
young stellar population. 
This is supported by very high equivalent widths EW(H$\beta$) 
of the H$\beta$ emission line with values of 200 -- 390\AA,
corresponding to a starburst age of $<$~3~Myr.

2. A strong [O {\sc iii}] $\lambda$4363\AA\ emission line is detected in all
objects, allowing element abundance determination by the direct
$T_{\rm e}$-method. We find low oxygen abundances 12 + log(O/H) in the range
7.53 -- 7.99. The Ne/O, S/O, Cl/O and Ar/O abundance ratios in all our galaxies
are similar to those found in low-metallicity blue compact dwarf (BCD) galaxies
\citep{I06}. On the other hand, the N/O and Fe/O abundance ratios in our CSFGs
at fixed 12+log(O/H) are higher by up to 0.5 dex and are almost independent
on the adopted ionisation-bounded, density-bounded or clumpy H~{\sc ii} models
which determine the ionisation correction factors.

3. Using six He~{\sc i} 3889, 4471, 5876, 6678, 7065, and 10830\AA\
emission lines and taking into account prosesses deviating their intensities
from the recombination values we derive the He mass fraction Y in
a self-consistent manner similar to that described e.g. by \citet{I14b}.
The derived Ys in our 11 galaxies are similar to that in the sample
of compact star-forming galaxies used by \citet{I14b} for the determination of
the primordial helium abundance.

4. The presence of high-excitation H~{\sc ii} regions with many strong permitted
and forbidden emission lines in a wide range of wavelengths and compact
structure of our galaxies with a single dominant H~{\sc ii} region results in
line profiles which can be fitted by single Gaussians in ten out of eleven
galaxies. This allows to study
velocity dispersions using many lines of various ions originating in different
parts of the H~{\sc ii} region. This approach differs from many other
similar spectroscopic studies where only velocity dispersions of brightest lines
H$\beta$, H$\alpha$ and [O~{\sc iii}] $\lambda$5007 were considered
\citep[e.g. ][]{Ch12,Ch14}. We find that $\sigma$(H$\beta$) increases with the
stellar mass and oxygen abundance and slightly decreases with
O$_{32}$. On the other hand, there is no correlation between $\sigma$(H$\beta$)
and EW(H$\beta$).

5. We compare velocity dispersions of H$\beta$ and various other emission lines.
We find that all other hydrogen lines in the optical and near-infrared ranges
have the same velocity dispersions as that of H$\beta$,
indicating that all these lines originate in the same volume of the H~{\sc ii}
region. We also note that [O~{\sc ii}] $\lambda$3726, $\lambda$3729,
[S~{\sc ii}] $\lambda$6717, $\lambda$6731
are narrower than H$\beta$, because they are produced in outer layers
of the H~{\sc ii} region. Contrary to that the velocity dispersion of the
He~{\sc ii} $\lambda$4686 emission line is considerably higher than that
of H$\beta$ indicating that this line is produced in the inner part of the
H~{\sc ii} region. The interesting appearance is for He~{\sc i} emission lines.
He~{\sc i} $\lambda$4471, $\lambda$5876 and $\lambda$7065 emission lines have
velocity dispersions similar to that of H$\beta$. The $\lambda$10830
emission line is somewhat broader (by $\sim$ 20\%), likely because
of radiative scattering and implying that it is emitted
in a somewhat different more dense region because of strong
dependence of its intensity on the electronic number density
compared to that for other He~{\sc i} emission lines.


\begin{acknowledgements} 
YII and NGG acknowledge support from the National Academy of
Sciences of Ukraine by its project no. 0126U000353.
The activities underlying the published results were carried out in the
framework of the project No. 224866 supported in the result of the Joint
Call `Ukrainian-Swiss Joint Research Projects: Call for Proposals 2023'.
RA acknowledges support of Grant
PID2023-147386NB-I00 funded by MICIU/AEI/10.13039/501100011033 and by
ERDF/EU, the Severo Ochoa award to the IAA CEX2021-001131-S.
\end{acknowledgements}

\input{ref1.tex}




\onecolumn
\begin{appendix}
\onecolumn

\section{Emission-line intensities, physical conditions and element abundances
of the ionised gas}
\input{tabintb1_1a.tex}
\setcounter{table}{0}
\input{tabintb1_1b.tex}

\setcounter{table}{0}
\input{tabintb1_2a.tex}

\setcounter{table}{0}
\input{tabintb1_2b.tex}
\setcounter{table}{1}
\setcounter{table}{1}
\input{tabcheb3_1a.tex}

\setcounter{table}{1}
\input{tabcheb3_2a.tex}

\section{Figures.}

\begin{figure}
\vspace{1.0cm}
\centering
\includegraphics[angle=0,width=0.70\linewidth]{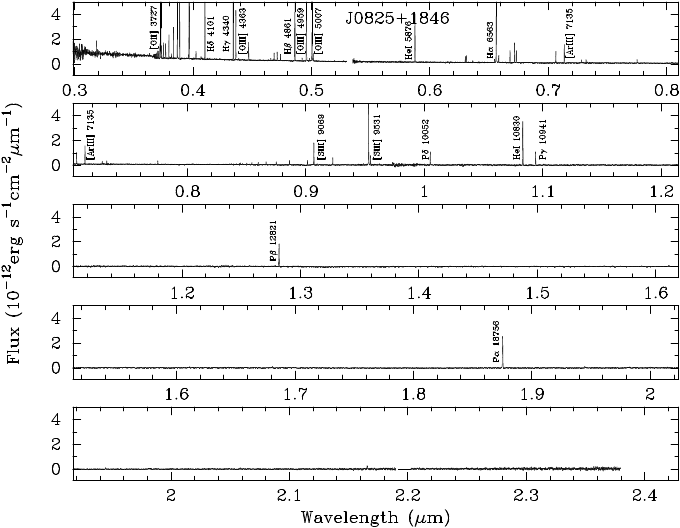}
\caption{The rest-frame spectrum of one of our galaxies, J0825+1846. }
\label{figb1}

\vspace{0.5cm}
\centering
\includegraphics[angle=0,width=0.70\linewidth]{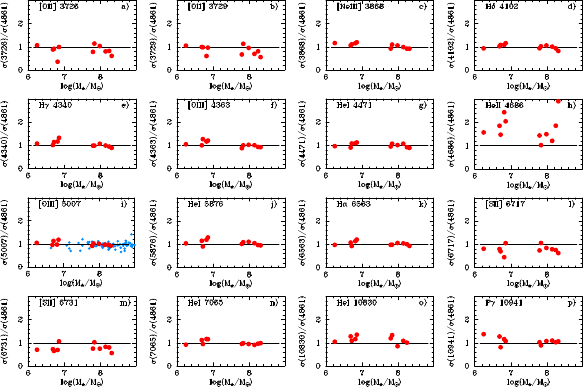}
\caption{The ratios of velocity dispersions of various lines to the velocity
dispersion of the H$\beta$ emission line in function of stellar mass.
Symbols are the same as in Fig.~\ref{fig9}.}
\label{figb2}
\end{figure}

\begin{figure}
\centering
\includegraphics[angle=0,width=0.70\linewidth]{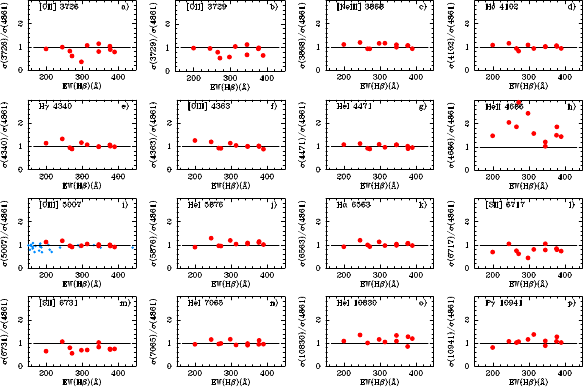}
\caption{The ratios of velocity dispersions of various lines to the velocity
dispersion of the H$\beta$ emission line in function of the H$\beta$
equivalent width. Symbols are the same as in Fig.~\ref{fig9}.}
\label{figb3}

\vspace{0.5cm}
\centering
\includegraphics[angle=0,width=0.70\linewidth]{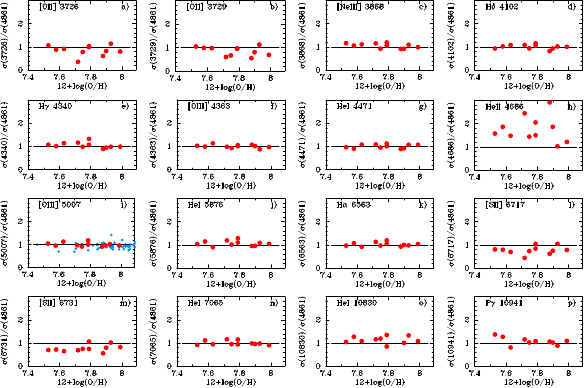}
\caption{The ratios of velocity dispersions of various lines to the velocity
dispersion of the H$\beta$ emission line in function of the oxygen abundance.
Symbols are the same as in Fig.~\ref{fig9}.}
\label{figb4}
\end{figure}

\end{appendix}
\end{document}

%% file: tab1.tex
  \begin{table*}
  \caption{General characteristics of galaxies
\label{tab1}}
\centering
\begin{tabular}{cccccccrcc} \hline
  \multicolumn{1}{c}{Name}&\multicolumn{1}{c}{R.A.(2000.0)}&\multicolumn{1}{c}{Dec.(2000.0)}&\multicolumn{1}{c}{Redshift}
&\multicolumn{1}{c}{$g$$^{\rm a}$}&\multicolumn{1}{c}{12+log(O/H)$^{\rm b}$}&\multicolumn{1}{c}{EW(H$\beta$)$^{\rm c}$}&\multicolumn{1}{c}{O$_{32}$}&\multicolumn{1}{c}{log($M_\star$)$^{\rm d}$}&log($\xi_{\rm ion}$)$^{\rm e}$  \\ \hline
J0007$+$0226& 00:07:24.48& $+$02:26:27.30& 0.06367& 19.73& 7.79&377& 39& 7.99&25.70\\ 
J0009$+$0234& 00:09:39.80& $+$02:34:19.69& 0.03921& 20.44& 7.53&314& 11& 6.25&25.67\\ 
J0159$+$0751& 01:59:52.75& $+$07:51:48.90& 0.06118& 19.47& 7.58&378& 42& 6.69&25.71\\ 
J0240$-$0828& 02:40:52.19& $-$08:28:27.41& 0.08222& 18.98& 7.93&346& 16& 7.84&25.60\\ 
J0825$+$1846& 08:25:40.44& $+$18:46:17.21& 0.03796& 18.94& 7.63&200& 11& 6.72&25.55\\ 
J1255$-$0213& 12:55:26.06& $-$02:13:34.13& 0.05188& 19.13& 7.79&245& 11& 6.86&25.60\\ 
J1310$+$0852& 13:10:54.30& $+$08:52:28.61& 0.04776& 18.97& 7.88&272& 19& 8.32&25.55\\ 
J1323$-$0132& 13:23:47.46& $-$01:32:52.00& 0.02246& 18.13& 7.72&298& 38& 6.82&25.62\\ 
J1325$+$0744& 13:25:04.04& $+$07:44:20.39& 0.07423& 19.89& 7.99&346& 17& 8.15&25.63\\ 
J1411$+$0550& 14:11:13.40& $+$05:50:35.05& 0.04940& 18.49& 7.90&266& 15& 8.25&25.64\\ 
J1526$+$2113& 15:26:54.45& $+$21:13:37.33& 0.05966& 19.50& 7.75&390& 28& 7.80&25.74\\ 
\hline
\end{tabular}

\tablefoot{$^{\rm a}$SDSS $g$-band magnitude. $^{\rm b}$Oxygen abundance derived by the direct $T_{\rm e}$-method from the Xshooter spectrum. $^{\rm c}$Rest-frame equivalent width
of the H$\beta$ emission line in the Xshooter spectrum, in \AA. $^{\rm d}$$M_\star$ is a stellar mass in M$_\odot$. $^{\rm e}$$\xi_{\rm ion}$ is a
production efficiency of ionizing photons in Hz erg$^{-1}$.}






\end{table*}

%% file: tab_obs.tex
  \begin{table*}
  \caption{Journal of observations.
\label{tab2}}
\centering
\begin{tabular}{cccccccccc} \hline
\multicolumn{1}{c}{Name}&\multicolumn{1}{c}{Date}&\multicolumn{3}{c}{Exposure (s)}&\multicolumn{1}{c}{PA}&\multicolumn{1}{c}{Seeing}&\multicolumn{1}{c}{Airmass}&\multicolumn{1}{c}{PSF in $r$-band (SDSS)}&\multicolumn{1}{c}{Standard star}\\ \cline{3-5}
                        &                        &   UVB   &   VIS   &    NIR      &(degree)&(arcsec)&            &(arcsec)              \\
\hline
J0007$+$0226&2021-09-28&2760&2880&500&341&1.03&1.2&1.18&Feige110\\
J0009$+$0234&2021-12-02&2760&2880&500&219&0.84&1.3&0.96&Feige110\\
J0159$+$0751&2021-11-07&2760&2880&500&216&0.96&1.4&1.11&Feige110\\
            &2021-11-09&2760&2880&500&210&0.83&1.3& ...&Feige110\\
J0240$-$0828&2021-11-07&1000&1000&160&321&0.81&1.2&1.22&Feige110\\
J0825$+$1846&2021-12-04&2760&2880&500&147&0.84&1.5&1.09&Feige110\\
J1255$-$0213&2022-05-26&2760&2880&500&213&0.76&1.1&1.36&GD153\\
J1310$+$0852&2022-05-02&2760&2880&500&323&0.44&1.3&1.21&LTT3218\\
J1323$-$0132&2022-05-29& 940& 940&160&337&0.89&1.1&1.32&LTT3218\\
J1325$+$0744&2022-05-04&2760&2880&500&331&0.95&1.2&1.20&LTT3218\\
J1411$+$0550&2022-05-02&2760&2880&500&328&1.22&1.2&1.03&LTT3218\\
            &2022-05-26&2760&2880&500&329&0.97&1.2& ...&GD153\\
            &2022-05-28&2760&2880&500&210&1.18&1.3& ...&LTT7987\\
J1526$+$2113&2022-04-02&2760&2880&500&316&1.06&1.4&1.14&LTT3218\\
\hline
\end{tabular}
\end{table*}

%% file: tab_cloudy.tex
  \begin{table}
  \caption{Input parameters of {\sc cloudy} density-bounded models.
\label{tab3}}
\centering
\begin{tabular}{lc} \hline
\multicolumn{1}{l}{Parameter}&\multicolumn{1}{c}{Value}\\ \hline
12+log(O/H)                        &7.6\\
log(C/O), log(N/O), log(Fe/O)      &--0.8, --1.6, --1.5\\
Ionization photon rate $Q$, s$^{-1}$&10$^{52}$, 10$^{53}$\\
Filling factor $f$                 & 0.01, 0.03, 0.1, 0.3, 1.0\\
Electron number density $N_{\rm e}$, cm$^{-3}$&10, 100, 200, 500\\
H~{\sc i} column density N(H~{\sc i}), cm$^{-2}$&10$^{17}$, 10$^{18}$, 10$^{19}$, 10$^{20}$\\
Starburst age, Myr& 2, 3\\
\hline
\end{tabular}
\end{table}

%% file: tab_minim.tex
  \begin{table}
  \caption{Range of parameteres for minimisation technique in Y determination.
\label{tab4}}
\centering
\begin{tabular}{lc} \hline
\multicolumn{1}{l}{Parameter}&\multicolumn{1}{c}{Value}\\ \hline
Electron temperature $T_{\rm e}$, K           &(0.95--1.05)$\times$$T_{\rm e}$(O~{\sc iii})\\
Electron number density $N_{\rm e}$, cm$^{-3}$  &10--1000\\
Optical depth $\tau$($\lambda$3889)            &0--5\\
\hline
\end{tabular}
\end{table}

%% file: tabhe2.tex
  \begin{table*}
  \caption{Helium mass fraction
\label{tab5}}
\centering
\begin{tabular}{cccccccccc} \hline
\multicolumn{1}{c}{Name}&\multicolumn{1}{c}{Y(mean of six lines)$^{\rm a}$}&\multicolumn{1}{c}{Y(3889)$^{\rm a}$}&\multicolumn{1}{c}{Y(4471)$^{\rm a}$}&\multicolumn{1}{c}{Y(5876)$^{\rm a}$} \\
                        &(Y(mean of five lines))$^{\rm b}$                  &\multicolumn{1}{c}{Y(6678)$^{\rm a}$}&\multicolumn{1}{c}{Y(7065)$^{\rm a}$}&\multicolumn{1}{c}{Y(10830)$^{\rm a}$} \\ \hline
J0007$+$0226&0.2566$\pm$0.0027  &0.2550$\pm$0.0090&0.2629$\pm$0.0102&0.2574$\pm$0.0050 \\ 
            &(0.2647$\pm$0.0032)&0.2660$\pm$0.0071&0.2557$\pm$0.0064&0.2571$\pm$0.0059 \\ \\
J0009$+$0234&0.2660$\pm$0.0036  &0.2866$\pm$0.0112&0.2735$\pm$0.0094&0.2936$\pm$0.0058 \\ 
            &(0.2808$\pm$0.0040)&0.3030$\pm$0.0147&0.2398$\pm$0.0098&0.2052$\pm$0.0084 \\ \\
J0159$+$0751&0.2688$\pm$0.0029  &0.2572$\pm$0.0074&0.2576$\pm$0.0066&0.2798$\pm$0.0057 \\ 
            &(0.2326$\pm$0.0028)&0.2762$\pm$0.0078&0.3598$\pm$0.0185&0.2667$\pm$0.0060 \\ \\
J0240$-$0828&0.2606$\pm$0.0029 &0.2667$\pm$0.0079&0.2538$\pm$0.0064&0.2647$\pm$0.0061 \\ 
            &(0.2550$\pm$0.0033)&0.2693$\pm$0.0170&0.2659$\pm$0.0074&0.2581$\pm$0.0056 \\ \\
J0825$+$1846&0.2798$\pm$0.0031 &0.2945$\pm$0.0086&0.2815$\pm$0.0064&0.2820$\pm$0.0135 \\ 
            &(0.2811$\pm$0.0036)&0.2609$\pm$0.0109&0.2817$\pm$0.0061&0.2738$\pm$0.0064 \\ \\
J1255$-$0213&0.2686$\pm$0.0024  &0.2932$\pm$0.0069&0.2594$\pm$0.0052&0.2645$\pm$0.0052 \\ 
            &(0.2591$\pm$0.0025)&0.2622$\pm$0.0063&0.2782$\pm$0.0064&0.2658$\pm$0.0060 \\ \\
J1310$+$0852&0.1314$\pm$0.0020 &0.3339$\pm$0.0093&0.2617$\pm$0.0073&0.2513$\pm$0.0075 \\ 
            &(0.2570$\pm$0.0034)&0.2471$\pm$0.0098&0.2103$\pm$0.0058&0.0748$\pm$0.0025 \\ \\
J1323$-$0132&0.2603$\pm$0.0027 &0.2598$\pm$0.0072&0.2659$\pm$0.0065&0.2601$\pm$0.0048 \\ 
            &(0.2703$\pm$0.0031)&0.2677$\pm$0.0072&0.2599$\pm$0.0109&0.2607$\pm$0.0060 \\ \\
J1325$+$0744&0.2508$\pm$0.0031  &0.2434$\pm$0.0089&0.2513$\pm$0.0153&0.2521$\pm$0.0050 \\ 
            &(0.2571$\pm$0.0037)&0.2794$\pm$0.0098&0.2451$\pm$0.0093&0.2526$\pm$0.0065 \\ \\
J1411$+$0550&0.2488$\pm$0.0020  &0.2507$\pm$0.0061&0.2507$\pm$0.0044&0.2524$\pm$0.0044 \\ 
            &(0.2473$\pm$0.0021)&0.2654$\pm$0.0047&0.2495$\pm$0.0044&0.2485$\pm$0.0052 \\ \\
J1526$+$2113&0.2634$\pm$0.0027  &0.2649$\pm$0.0071&0.2623$\pm$0.0065&0.2629$\pm$0.0050 \\ 
            &(0.2595$\pm$0.0030)&0.2780$\pm$0.0066&0.2801$\pm$0.0228&0.2631$\pm$0.0059 \\
\hline
\end{tabular}

\tablefoot{$^{\rm a}$Derived from minimisation of the abundance differences of six He~{\sc i} emission lines, He~{\sc i} $\lambda$3889, 4471, 5876, 6678, 7065 and
10830. $^{\bf b}$Derived from minimisation of the abundance differences of five He~{\sc i} emission lines excluding He~{\sc i} $\lambda$10830 emission line.}

\end{table*}

%% file: tab_broad1_6.tex
\begin{table*}
\caption{Velocity dispersions $\sigma$ in galaxies from this paper in km s$^{-1}$} \label{tab6}
\begin{tabular}{cccccccccc} \hline
            &\multicolumn{9}{c}{Emission line}\\ \cline{2-10}
Name        &[O~{\sc ii}]&[O~{\sc ii}]&[Ne~{\sc iii}]&H$\delta$ &H$\gamma$ &[O~{\sc iii}]&He~{\sc i}&He~{\sc ii}&H$\beta$  \\
            &   3726$\AA$&   3729$\AA$&   3868$\AA$&   4102$\AA$&   4340$\AA$&   4363$\AA$&   4471$\AA$&   4686$\AA$&   4861$\AA$\\ \hline
J0007$+$0226&28.4$\pm$0.4&26.2$\pm$0.4&29.1$\pm$0.1&29.1$\pm$0.1&29.4$\pm$0.1&28.2$\pm$0.2&27.8$\pm$0.7&41.3$\pm$1.9&27.7$\pm$0.1\\
J0009$+$0234&20.0$\pm$0.3&19.7$\pm$0.3&21.8$\pm$0.2&17.5$\pm$0.3&20.2$\pm$0.2&19.6$\pm$0.4&18.1$\pm$0.7&29.6$\pm$2.6&18.9$\pm$0.1\\
J0159$+$0751&17.2$\pm$0.4&19.3$\pm$0.3&20.6$\pm$0.1&19.9$\pm$0.1&19.7$\pm$0.1&19.4$\pm$0.1&17.6$\pm$0.5&36.3$\pm$1.6&19.6$\pm$0.1\\
J0240$-$0828&36.0$\pm$0.3&35.7$\pm$0.3&34.6$\pm$0.1&32.2$\pm$0.1&31.2$\pm$0.1&31.5$\pm$0.2&33.9$\pm$0.7&32.4$\pm$0.8&31.8$\pm$0.1\\
J0825$+$1846&21.1$\pm$0.2&22.6$\pm$0.2&25.5$\pm$0.1&24.9$\pm$0.2&26.3$\pm$0.2&29.0$\pm$0.4&24.9$\pm$0.5&34.1$\pm$1.0&23.2$\pm$0.1\\
J1255$-$0213&18.6$\pm$0.1&18.1$\pm$0.1&22.4$\pm$0.1&21.7$\pm$0.1&24.9$\pm$0.1&22.5$\pm$0.1&21.1$\pm$0.2&38.6$\pm$1.0&18.9$\pm$0.1\\
J1310$+$0852&15.9$\pm$0.2&14.5$\pm$0.2&24.0$\pm$0.2&21.4$\pm$0.2&23.3$\pm$0.1&24.0$\pm$0.2&23.3$\pm$0.4&76.4$\pm$9.5&26.3$\pm$0.1\\
J1323$-$0132&~~5.8$\pm$0.2&~~9.8$\pm$0.2&18.8$\pm$0.1&17.7$\pm$0.1&19.0$\pm$0.1&18.6$\pm$0.1&17.8$\pm$0.3&39.9$\pm$4.3&16.4$\pm$0.1\\
J1325$+$0744&24.9$\pm$0.4&21.6$\pm$0.3&30.7$\pm$0.1&31.3$\pm$0.2&30.9$\pm$0.1&31.3$\pm$0.2&33.3$\pm$1.4&37.7$\pm$7.1&31.3$\pm$0.1\\
J1411$+$0550&28.6$\pm$0.1&28.1$\pm$0.1&32.1$\pm$0.1&32.5$\pm$0.1&33.0$\pm$0.1&32.4$\pm$0.1&32.1$\pm$0.4&65.1$\pm$1.8&35.1$\pm$0.1\\
J1526$+$2113&19.9$\pm$0.2&17.0$\pm$0.1&23.5$\pm$0.1&23.8$\pm$0.1&25.1$\pm$0.1&22.5$\pm$0.1&24.2$\pm$0.8&36.7$\pm$1.4&25.6$\pm$0.1\\ \hline \\ \cline{2-9}

            &\multicolumn{8}{c}{Emission line}\\ \cline{2-9}
Name        &[O~{\sc iii}] &He~{\sc i}  &H$\alpha$ &[S~{\sc ii}]&[S~{\sc ii}]&He~{\sc i} &He~{\sc i}&P$\gamma$\\
            &   5007$\AA$&   5876$\AA$&   6563$\AA$&   6717$\AA$&   6731$\AA$&   7065$\AA$&  10830$\AA$&  10941$\AA$\\ \cline{2-9}
J0007$+$0226&27.7$\pm$0.1&30.5$\pm$0.3&29.1$\pm$0.1&23.2$\pm$0.4&20.6$\pm$0.4&26.1$\pm$0.3&23.5$\pm$0.4&29.6$\pm$1.5\\
J0009$+$0234&19.6$\pm$0.1&19.5$\pm$0.2&18.3$\pm$0.1&15.2$\pm$0.3&13.2$\pm$0.3&17.3$\pm$0.3&19.7$\pm$2.4&25.8$\pm$4.5\\
J0159$+$0751&18.6$\pm$0.1&22.4$\pm$0.2&20.9$\pm$0.1&15.5$\pm$0.2&14.1$\pm$0.2&21.8$\pm$0.2&24.9$\pm$0.2&24.8$\pm$0.7 \\
J0240$-$0828&32.2$\pm$0.1&34.4$\pm$0.2&31.1$\pm$0.1&33.3$\pm$0.4&32.4$\pm$0.4&31.0$\pm$0.2&42.1$\pm$0.4&28.1$\pm$1.2\\
J0825$+$1846&26.0$\pm$0.1&20.9$\pm$0.1&21.3$\pm$0.1&16.0$\pm$0.2&15.0$\pm$0.2&21.9$\pm$0.3&25.1$\pm$0.4&18.7$\pm$1.3\\
J1255$-$0213&22.2$\pm$0.1&24.3$\pm$0.3&22.6$\pm$0.1&19.7$\pm$0.3&20.0$\pm$0.3&21.7$\pm$0.3&25.4$\pm$0.3&20.2$\pm$0.5\\
J1310$+$0852&23.9$\pm$0.1&24.9$\pm$0.3&24.3$\pm$0.1&16.2$\pm$0.2&14.6$\pm$0.2&25.8$\pm$0.5&     ...    &27.8$\pm$2.6\\
J1323$-$0132&15.8$\pm$0.1&19.5$\pm$0.2&18.5$\pm$0.1& ~~7.2$\pm$0.5&11.3$\pm$0.5&19.0$\pm$0.3&18.9$\pm$0.3&18.9$\pm$0.9\\
J1325$+$0744&29.8$\pm$0.1&32.6$\pm$0.3&32.3$\pm$0.1&24.4$\pm$0.3&25.8$\pm$0.4&28.3$\pm$0.6&33.8$\pm$0.6&34.1$\pm$3.5\\
J1411$+$0550&33.7$\pm$0.1&33.7$\pm$0.2&35.0$\pm$0.1&26.2$\pm$0.3&27.9$\pm$0.3&33.6$\pm$0.2&35.2$\pm$0.3&35.7$\pm$0.8\\
J1526$+$2113&23.2$\pm$0.1&25.9$\pm$0.2&24.8$\pm$0.1&18.7$\pm$0.3&19.1$\pm$0.3&24.3$\pm$0.3&30.4$\pm$0.3&26.1$\pm$1.0\\ \cline{2-9}
\end{tabular}
\end{table*}

%% file: tab10830.tex
  \begin{table}
  \caption{Peak separation of the He~{\sc i} $\lambda$10830\AA\ emission line
\label{tab7}}
\centering
\begin{tabular}{ccccc} \hline
\multicolumn{1}{c}{Name}&\multicolumn{1}{c}{Separation}&&\multicolumn{1}{c}{Name}&\multicolumn{1}{c}{Separation} \\
&\multicolumn{1}{c}{km s$^{-1}$}&&&\multicolumn{1}{c}{km s$^{-1}$} \\ \hline
J0007$+$0226& 92.2$\pm$2.8&&J1310$+$0852& ... \\
J0009$+$0234&      ...    &&J1323$-$0132& 56.8$\pm$3.3 \\
J0159$+$0751& 81.2$\pm$3.3&&J1325$+$0744& 92.2$\pm$9.7 \\
J0240$-$0828&      ...    &&J1411$+$0550& 99.7$\pm$6.2 \\
J0825$+$1846&      ...    &&J1255$-$0213& 64.0$\pm$4.7 \\
J1526$+$2113&130.2$\pm$7.2 \\
\hline
\end{tabular}


\end{table}

%% file: tabintb1_1a.tex
\begin{table}[t]
\caption{Emission-line intensities}
\label{taba1}
\begin{tabular}{lrrrrrrr} \hline
Line             & \multicolumn{1}{c}{$\lambda$}& \multicolumn{6}{c}{100$\times$$I(\lambda)$/$I$(H$\beta$)$^{\rm a}$}\\ \cline{3-8}
& & \multicolumn{6}{c}{Galaxy} \\ \cline{3-8}
 & & J0007$+$0226&J0009$+$0234&J0159$+$0751&J0240$-$0828&J0825$+$1846&J1255$-$0213 \\ \hline
$[$Fe~{\sc iv}$]$    & 3096.00&...~~~~~     &...~~~~~     &...~~~~~     &  1.8$\pm$0.4&...~~~~~     &  1.8$\pm$0.3\\
O~{\sc iii}          & 3132.79&...~~~~~     &...~~~~~     &  2.0$\pm$0.1&  2.0$\pm$0.3&...~~~~~     &...~~~~~     \\
He~{\sc i}           & 3187.74&  3.4$\pm$0.6&  4.0$\pm$0.8&  3.1$\pm$0.1&  2.9$\pm$0.3&  4.8$\pm$0.7&  4.8$\pm$1.3\\
He~{\sc ii}          & 3203.10&  0.6$\pm$0.4&...~~~~~     &  1.0$\pm$0.3&  1.3$\pm$0.3&  1.3$\pm$0.5&  0.8$\pm$0.2\\
$[$Ne~{\sc v}$]$     & 3426.85&...~~~~~     &  0.8$\pm$0.2&...~~~~~     &  0.4$\pm$0.1&...~~~~~     &...~~~~~     \\
O~{\sc iii}          & 3444.07&...~~~~~     &...~~~~~     &...~~~~~     &  0.4$\pm$0.1&...~~~~~     &...~~~~~     \\
He~{\sc i}           & 3587.28&...~~~~~     &...~~~~~     &...~~~~~     &  0.4$\pm$0.1&...~~~~~     &...~~~~~     \\
He~{\sc i}           & 3634.25&  0.7$\pm$0.3&...~~~~~     &  0.6$\pm$0.1&  0.4$\pm$0.1&  0.6$\pm$0.1&...~~~~~     \\
H22                  & 3676.37&  0.5$\pm$0.2&...~~~~~     &  0.8$\pm$0.1&  0.5$\pm$0.1&  0.9$\pm$0.2&  0.8$\pm$0.1\\
H21                  & 3679.31&  1.0$\pm$0.2&...~~~~~     &  0.8$\pm$0.1&  0.5$\pm$0.1&  1.0$\pm$0.2&  1.1$\pm$0.1\\
H20                  & 3682.81&  1.2$\pm$0.2&...~~~~~     &  0.7$\pm$0.1&  0.7$\pm$0.1&  1.3$\pm$0.2&  1.1$\pm$0.1\\
H19                  & 3686.83&  0.9$\pm$0.2&...~~~~~     &  1.0$\pm$0.1&  0.7$\pm$0.1&  1.2$\pm$0.2&  1.2$\pm$0.1\\
H18                  & 3691.55&  1.4$\pm$0.2&  1.3$\pm$0.3&  1.2$\pm$0.1&  0.8$\pm$0.1&  1.2$\pm$0.2&  1.3$\pm$0.1\\
H17                  & 3697.15&  1.5$\pm$0.2&  1.3$\pm$0.2&  1.2$\pm$0.1&  1.0$\pm$0.1&  1.4$\pm$0.3&  1.4$\pm$0.1\\
H16+O~{\sc iii}      & 3703.30&  1.9$\pm$0.2&  1.8$\pm$0.2&  1.9$\pm$0.1&  1.8$\pm$0.1&  2.5$\pm$0.3&  2.5$\pm$0.1\\
H15                  & 3711.97&  1.8$\pm$0.3&  3.4$\pm$0.4&  3.1$\pm$0.3&  3.1$\pm$0.3&  3.0$\pm$0.9&  4.4$\pm$0.3\\
H14                  & 3721.94&  2.6$\pm$0.2&  3.2$\pm$0.2&  2.3$\pm$0.1&  2.8$\pm$0.2&  3.1$\pm$0.2&  3.6$\pm$0.1\\
$[$O~{\sc ii}$]$     & 3726.03&  9.1$\pm$0.4& 17.6$\pm$0.6&  7.0$\pm$0.3& 23.9$\pm$0.8& 21.7$\pm$0.7& 25.2$\pm$0.8\\
$[$O~{\sc ii}$]$     & 3728.82& 11.5$\pm$0.4& 24.6$\pm$0.8&  8.5$\pm$0.3& 21.7$\pm$0.7& 26.6$\pm$0.9& 34.2$\pm$1.1\\
H13                  & 3734.37&  2.7$\pm$0.3&  2.8$\pm$0.2&  2.6$\pm$0.1&  2.2$\pm$0.2&  2.8$\pm$0.2&  2.8$\pm$0.1\\
H12                  & 3750.15&  3.6$\pm$0.3&  4.9$\pm$0.4&  4.5$\pm$0.3&  4.2$\pm$0.5&  4.9$\pm$0.3&  6.0$\pm$0.3\\
H11                  & 3770.63&  4.2$\pm$0.4&  6.1$\pm$0.4&  5.4$\pm$0.3&  5.4$\pm$0.3&  5.6$\pm$0.3&  6.6$\pm$0.3\\
H10                  & 3797.90&  5.7$\pm$0.5&  6.9$\pm$0.3&  6.9$\pm$0.4&  6.7$\pm$0.3&  7.0$\pm$0.3&  8.3$\pm$0.3\\
He~{\sc i}           & 3819.64&  1.2$\pm$0.2&  1.2$\pm$0.1&  1.2$\pm$0.3&  1.1$\pm$0.1&  1.0$\pm$0.1&  1.2$\pm$0.1\\
H9                   & 3835.39&  7.6$\pm$0.4&  8.6$\pm$0.4& 10.6$\pm$0.4&  8.4$\pm$0.4&  9.4$\pm$0.4& 10.4$\pm$0.4\\
$[$Ne~{\sc iii}$]$   & 3868.76& 62.0$\pm$2.0& 36.4$\pm$1.2& 52.9$\pm$1.7& 46.8$\pm$1.8& 38.7$\pm$1.2& 54.7$\pm$1.7\\
He~{\sc i}+H8        & 3889.00& 18.9$\pm$0.7& 21.9$\pm$0.9& 18.3$\pm$0.6& 19.3$\pm$0.7& 22.9$\pm$0.8& 22.4$\pm$0.7\\
He~{\sc i}           & 3964.73&  0.6$\pm$0.2&  0.7$\pm$0.1&  0.8$\pm$0.1&  1.3$\pm$0.1&  0.9$\pm$0.1&  0.7$\pm$0.1\\
$[$Ne~{\sc iii}$]$   & 3967.14& 18.1$\pm$0.6& 14.0$\pm$0.5& 17.1$\pm$0.6& 15.1$\pm$0.5& 16.3$\pm$0.5& 18.8$\pm$0.6\\
H7                   & 3970.07& 15.1$\pm$0.5& 16.7$\pm$0.5& 15.3$\pm$0.5& 14.3$\pm$0.5& 15.3$\pm$0.5& 17.0$\pm$0.5\\
He~{\sc i}           & 4009.26&...~~~~~     &...~~~~~     &...~~~~~     &...~~~~~     &  0.3$\pm$0.1&...~~~~~     \\
He~{\sc i}           & 4026.19&  1.4$\pm$0.4&  2.1$\pm$0.1&  1.9$\pm$0.2&  1.8$\pm$0.1&  1.9$\pm$0.1&  1.9$\pm$0.1\\
$[$S~{\sc ii}$]$     & 4068.60&...~~~~~     &  0.9$\pm$0.2&  0.3$\pm$0.1&  0.8$\pm$0.1&  0.8$\pm$0.1&  0.9$\pm$0.1\\
$[$S~{\sc ii}$]$     & 4076.35&...~~~~~     &...~~~~~     &...~~~~~     &  0.2$\pm$0.1&  0.5$\pm$0.1&...~~~~~     \\
H$\delta$            & 4101.74& 26.4$\pm$0.9& 24.4$\pm$0.8& 26.8$\pm$0.8& 26.8$\pm$0.8& 27.3$\pm$0.9& 29.7$\pm$0.9\\
He~{\sc i}           & 4120.84&  0.4$\pm$0.1&...~~~~~     &  0.3$\pm$0.1&  0.2$\pm$0.1&  0.3$\pm$0.1&  0.3$\pm$0.1\\
He~{\sc i}           & 4143.76&...~~~~~     &...~~~~~     &  0.5$\pm$0.1&  0.4$\pm$0.1&  0.3$\pm$0.1&  0.4$\pm$0.1\\
$[$Fe~{\sc v}$]$     & 4227.20&...~~~~~     &  0.3$\pm$0.1&  1.3$\pm$0.1&  0.4$\pm$0.1&  0.4$\pm$0.1&  0.3$\pm$0.1\\
$[$Fe~{\sc ii}$]$    & 4287.33&...~~~~~     &...~~~~~     &...~~~~~     &  0.3$\pm$0.1&...~~~~~     &  0.2$\pm$0.1\\
H$\gamma$            & 4340.47& 48.3$\pm$1.4& 46.9$\pm$1.4& 47.1$\pm$1.4& 47.1$\pm$1.4& 48.1$\pm$1.4& 43.1$\pm$1.3\\
$[$O~{\sc iii}$]$    & 4363.21& 21.5$\pm$0.7& 15.2$\pm$0.5& 21.9$\pm$0.7& 14.5$\pm$0.4& 14.6$\pm$0.5& 15.0$\pm$0.5\\
He~{\sc i}           & 4387.93&  0.6$\pm$0.1&  0.7$\pm$0.1&  0.6$\pm$0.1&  0.6$\pm$0.1&  0.4$\pm$0.1&  0.5$\pm$0.1\\
He~{\sc i}           & 4471.48&  4.1$\pm$0.2&  4.1$\pm$0.2&  3.9$\pm$0.1&  4.0$\pm$0.1&  4.0$\pm$0.1&  3.8$\pm$0.1\\
$[$Fe~{\sc iii}$]$   & 4658.10&...~~~~~     &...~~~~~     &  0.3$\pm$0.1&  0.7$\pm$0.1&  0.7$\pm$0.1&  0.6$\pm$0.1\\
He~{\sc ii}          & 4686.00&  1.6$\pm$0.1&  1.8$\pm$0.1&  1.4$\pm$0.1&  2.8$\pm$0.1&  2.4$\pm$0.1&  1.9$\pm$0.1\\
$[$Ar~{\sc iv}$]$    & 4711.37&  5.2$\pm$0.2&  2.5$\pm$0.1&  0.8$\pm$0.1&  1.6$\pm$0.1&  2.2$\pm$0.1&  1.9$\pm$0.1\\
He~{\sc i}           & 4713.14&...~~~~~     &  0.7$\pm$0.1&  0.8$\pm$0.1&  0.6$\pm$0.1&  0.6$\pm$0.1&  0.5$\pm$0.1\\
$[$Ar~{\sc iv}$]$    & 4740.20&  3.3$\pm$0.1&  2.0$\pm$0.1&  3.2$\pm$0.1&  1.2$\pm$0.1&  1.7$\pm$0.1&  1.6$\pm$0.1\\
H$\beta$             & 4861.33&100.0$\pm$2.9&100.0$\pm$2.9&100.0$\pm$2.9&100.0$\pm$2.9&100.0$\pm$2.9&100.0$\pm$2.9\\
He~{\sc i}           & 4921.93&  1.3$\pm$0.1&  1.1$\pm$0.1&  1.2$\pm$0.1&  1.1$\pm$0.1&  0.9$\pm$0.1&  1.2$\pm$0.1\\
$[$O~{\sc iii}$]$    & 4958.92&265.5$\pm$7.6&166.6$\pm$4.8&214.3$\pm$6.1&240.6$\pm$6.9&174.8$\pm$5.0&212.2$\pm$6.1\\
$[$Fe~{\sc iii}$]$   & 4988.00&...~~~~~     &  0.7$\pm$0.1&...~~~~~     &  0.5$\pm$0.1&  0.8$\pm$0.1&  0.8$\pm$0.1\\
$[$O~{\sc iii}$]$    & 5006.85&794.8$\pm$23.&478.6$\pm$13.&644.4$\pm$18.&723.5$\pm$21.&522.8$\pm$15.&630.0$\pm$18.\\
He~{\sc i}           & 5015.68&  2.8$\pm$0.1&  2.4$\pm$0.1&  1.7$\pm$0.1&  1.1$\pm$0.1&  2.4$\pm$0.1&  1.8$\pm$0.1\\
Si~{\sc ii}          & 5041.03&...~~~~~     &...~~~~~     &...~~~~~     &  0.3$\pm$0.1&  0.3$\pm$0.1&...~~~~~     \\
He~{\sc i}           & 5047.80&...~~~~~     &...~~~~~     &  0.3$\pm$0.1&...~~~~~     &  0.3$\pm$0.1&  0.2$\pm$0.1\\
$[$Fe~{\sc iii}$]$   & 5270.16&...~~~~~     & ...~~~~~    & ...~~~~~    & ...~~~~~    &  0.3$\pm$0.1&...~~~~~     \\
He~{\sc i}           & 5875.60& 10.5$\pm$0.3& 11.5$\pm$0.4& 11.3$\pm$0.4& 12.2$\pm$0.4& 11.0$\pm$0.6& 10.4$\pm$0.3\\
$[$O~{\sc i}$]$      & 6300.30&  0.8$\pm$0.1&  1.3$\pm$0.1&  0.8$\pm$0.1&  1.7$\pm$0.1&  1.3$\pm$0.1&  1.3$\pm$0.1\\
$[$S~{\sc iii}$]$    & 6312.10&  1.1$\pm$0.1&  1.4$\pm$0.1&  0.4$\pm$0.1&  1.1$\pm$0.1&  1.5$\pm$0.1&  1.6$\pm$0.1\\
$[$O~{\sc i}$]$      & 6363.80&  0.6$\pm$0.1&  0.6$\pm$0.1&...~~~~~     &  0.5$\pm$0.1&  0.4$\pm$0.1&  0.5$\pm$0.1\\
\hline
\end{tabular}
\end{table}

%% file: tabintb1_1b.tex
\begin{table}[t]
\caption{continued}
\begin{tabular}{lrrrrrrr} \hline
Line             & \multicolumn{1}{c}{$\lambda$}& \multicolumn{6}{c}{100$\times$$I(\lambda)$/$I$(H$\beta$)$^{\rm a}$}\\ \cline{3-8}
& & \multicolumn{6}{c}{Galaxy} \\ \cline{3-8}
 & & J0007$+$0226&J0009$+$0234&J0159$+$0751&J0240$-$0828&J0825$+$1846&J1255$-$0213 \\ \hline
$[$N~{\sc ii}$]$     & 6548.10&  0.3$\pm$0.1&  0.9$\pm$0.1&  0.5$\pm$0.1&  3.1$\pm$0.1&  0.8$\pm$0.1&  1.0$\pm$0.1\\
H$\alpha$            & 6562.80&277.0$\pm$8.6&275.1$\pm$0.6&275.2$\pm$8.5&280.1$\pm$8.7&266.9$\pm$8.3&279.5$\pm$8.7\\
$[$N~{\sc ii}$]$     & 6583.40&  1.4$\pm$0.1&  2.5$\pm$0.1&  1.2$\pm$0.1&  8.5$\pm$0.3&  2.2$\pm$0.1&  2.7$\pm$0.1\\
He~{\sc i}           & 6678.10&  2.9$\pm$0.1&  3.3$\pm$0.2&  2.9$\pm$0.1&  3.0$\pm$0.2&  2.8$\pm$0.1&  2.9$\pm$0.1\\
$[$S~{\sc ii}$]$     & 6716.40&  2.4$\pm$0.1&  4.6$\pm$0.2&  1.5$\pm$0.1&  4.3$\pm$0.2&  4.8$\pm$0.2&  5.5$\pm$0.2\\
$[$S~{\sc ii}$]$     & 6730.80&  1.7$\pm$0.1&  3.5$\pm$0.1&  1.2$\pm$0.1&  4.7$\pm$0.2&  3.5$\pm$0.1&  4.2$\pm$0.1\\
He~{\sc i}           & 7065.30&  3.6$\pm$0.1&  5.0$\pm$0.2&  7.0$\pm$0.4&  5.6$\pm$0.2&  3.1$\pm$0.1&  2.7$\pm$0.1\\
$[$Ar~{\sc iii}$]$   & 7135.80&  4.1$\pm$0.2&  4.2$\pm$0.2&  2.0$\pm$0.1&  4.6$\pm$0.2&  4.4$\pm$0.2&  5.3$\pm$0.2\\
$[$Ar~{\sc iv}$]$    & 7170.62&...~~~~~     &...~~~~~     &...~~~~~     &...~~~~~     &  0.1$\pm$0.1&...~~~~~     \\
$[$Ar~{\sc iv}$]$    & 7262.76&  0.3$\pm$0.1&...~~~~~     &  0.2$\pm$0.1&...~~~~~     &  0.2$\pm$0.1&...~~~~~     \\
He~{\sc i}           & 7281.35&  0.7$\pm$0.1&  1.0$\pm$0.1&  0.8$\pm$0.1&  0.8$\pm$0.1&  0.7$\pm$0.1&  0.5$\pm$0.1\\
$[$O~{\sc ii}$]$     & 7319.90&  0.5$\pm$0.1&  1.2$\pm$0.4&  0.6$\pm$0.1&  1.6$\pm$0.1&  1.0$\pm$0.1&  1.0$\pm$0.1\\
$[$O~{\sc ii}$]$     & 7330.20&  0.4$\pm$0.1&  0.9$\pm$0.1&  0.5$\pm$0.1&  1.3$\pm$0.1&  0.6$\pm$0.1&  0.9$\pm$0.1\\
$[$Ar~{\sc iii}$]$   & 7751.12&  1.1$\pm$0.1&  1.2$\pm$0.1&  0.4$\pm$0.1&  0.9$\pm$0.1&  1.1$\pm$0.1&  1.4$\pm$0.1\\
$[$Cl~{\sc iv}$]$    & 8045.63&  0.4$\pm$0.1&...~~~~~     &  0.3$\pm$0.1&...~~~~~     &...~~~~~     &  0.1$\pm$0.1\\
P20                  & 8392.40& ...~~~~~    &...~~~~~     &...~~~~~     &...~~~~~     &  0.3$\pm$0.1&...~~~~~     \\
P19                  & 8413.32& ...~~~~~    &...~~~~~     &  0.3$\pm$0.1&...~~~~~     &  0.3$\pm$0.1&  0.4$\pm$0.1\\
P18                  & 8437.96&  0.3$\pm$0.1&...~~~~~     &  0.4$\pm$0.1&  0.3$\pm$0.1&  0.4$\pm$0.1&  0.4$\pm$0.1\\
O~{\sc i}            & 8446.34& ...~~~~~    &...~~~~~     &  0.3$\pm$0.1&  0.9$\pm$0.1&  0.6$\pm$0.1&...~~~~~     \\
P17                  & 8467.26&  0.4$\pm$0.1&...~~~~~     &  0.3$\pm$0.1&  0.4$\pm$0.1&  0.5$\pm$0.1&  0.4$\pm$0.1\\
P16                  & 8502.49&  0.6$\pm$0.1&...~~~~~     &  0.6$\pm$0.1&  0.4$\pm$0.1&  0.6$\pm$0.1&  0.7$\pm$0.1\\
P15                  & 8545.38&  0.8$\pm$0.1&  0.8$\pm$0.1&  0.6$\pm$0.1&  0.5$\pm$0.1&  0.7$\pm$0.1&...~~~~~     \\
P14                  & 8598.39&  0.7$\pm$0.1&  1.1$\pm$0.1&  0.9$\pm$0.1&...~~~~~     &  0.8$\pm$0.1&  0.7$\pm$0.1\\
P13                  & 8665.02&  0.9$\pm$0.1&  0.8$\pm$0.1&  0.9$\pm$0.1&...~~~~~     &  0.8$\pm$0.1&  0.8$\pm$0.1\\
P12                  & 8750.47&  1.3$\pm$0.1&  1.5$\pm$0.1&  0.5$\pm$0.1&  1.5$\pm$0.1&  1.1$\pm$0.1&  1.3$\pm$0.2\\
P11                  & 8862.79&  1.5$\pm$0.1&  2.0$\pm$0.1&  0.8$\pm$0.1&  1.3$\pm$0.1&  1.7$\pm$0.2&  2.4$\pm$0.2\\
P10                  & 9014.91&  2.0$\pm$0.1&...~~~~~     &  1.7$\pm$0.1&  1.9$\pm$0.2&  1.2$\pm$0.1&  2.7$\pm$0.1\\
$[$S~{\sc iii}$]$    & 9069.00&  6.2$\pm$0.3&  4.4$\pm$0.2&  3.1$\pm$0.1&  6.2$\pm$0.4&  7.1$\pm$0.3& 11.2$\pm$0.4\\
P9                   & 9229.02&  2.8$\pm$0.2&  1.7$\pm$0.1&  1.7$\pm$0.1&  2.1$\pm$0.2&  2.2$\pm$0.1&  2.5$\pm$0.1\\
$[$S~{\sc iii}$]$    & 9530.60& 16.6$\pm$0.7& 11.1$\pm$0.5&  6.5$\pm$0.4& 15.3$\pm$0.7& 22.1$\pm$0.1& 24.0$\pm$0.9\\
P8                   & 9545.98&  5.8$\pm$0.6&  1.9$\pm$0.2&  4.4$\pm$0.4&  3.0$\pm$0.2&  3.8$\pm$0.2&  3.6$\pm$0.3\\
P$\delta$            &10052.15&  6.7$\pm$0.4&  6.5$\pm$0.4&  5.4$\pm$0.2&  5.1$\pm$0.2&  5.6$\pm$0.3&  4.5$\pm$0.2\\
He~{\sc i}           &10830.00& 29.8$\pm$1.2& 13.5$\pm$0.7& 28.7$\pm$1.1& 57.3$\pm$2.3& 22.1$\pm$0.9& 21.3$\pm$0.9\\
P$\gamma$            &10941.12&  9.5$\pm$0.4&  7.4$\pm$0.4&  7.1$\pm$0.3&  7.4$\pm$0.3&  5.6$\pm$0.3&  7.2$\pm$0.3\\
P$\beta$             &12821.62&  8.4$\pm$0.4& 19.5$\pm$0.9&  1.7$\pm$0.1&  4.3$\pm$0.2& 13.1$\pm$0.6& 10.9$\pm$0.5\\
Br15                 &15704.96&...~~~~~     &...~~~~~     &...~~~~~     &...~~~~~     &...~~~~~     &  0.3$\pm$0.1\\
Br14                 &15884.88&...~~~~~     &...~~~~~     &  0.4$\pm$0.1&...~~~~~     &...~~~~~     &...~~~~~     \\
Br12                 &16411.68&  0.7$\pm$0.1&...~~~~~     &...~~~~~     &...~~~~~     &...~~~~~     &  0.8$\pm$0.1\\
Br11                 &16811.12&...~~~~~     &...~~~~~     &  0.7$\pm$0.1&  0.6$\pm$0.1&  0.8$\pm$0.1&  0.9$\pm$0.1\\
Br10                 &17366.86&...~~~~~     &...~~~~~     &...~~~~~     &  0.9$\pm$0.1&...~~~~~     &  0.5$\pm$0.1\\
Br9                  &18179.09&  1.0$\pm$0.1&...~~~~~     &...~~~~~     &  1.0$\pm$0.1&...~~~~~     &  0.8$\pm$0.1\\
P$\alpha$            &18756.20& 27.6$\pm$1.2& 18.9$\pm$0.9& 20.0$\pm$0.9& 25.7$\pm$1.1& 22.0$\pm$1.0& 21.9$\pm$1.0\\
Br$\delta$           &19450.88&  1.9$\pm$0.1&...~~~~~     &  1.8$\pm$0.1&  1.8$\pm$0.1&  1.0$\pm$0.1&  1.7$\pm$0.1\\
He~{\sc i}           &20580.00&...~~~~~     &...~~~~~     &...~~~~~     &  0.5$\pm$0.1&  0.8$\pm$0.1&...~~~~~     \\
Br$\gamma$           &21661.21&...~~~~~     &...~~~~~     &...~~~~~     &  1.9$\pm$0.1&  2.5$\pm$0.1&...~~~~~     \\ \\
$C$(H$\beta$)$^{\rm b}$&    &0.085$\pm$0.037&0.070$\pm$0.037&0.145$\pm$0.037&0.085$\pm$0.037&0.000$\pm$0.037&0.170$\pm$0.037\\
$EW$(H$\beta$)$^{\rm c}$&        &377.3$\pm$1.3&314.0$\pm$0.9&378.2$\pm$0.7&345.9$\pm$0.5&200.2$\pm$0.3&244.5$\pm$0.4\\
$F$(H$\beta$)$^{\rm d}$&       &24.42$\pm$0.08&16.01$\pm$0.05&29.28$\pm$0.06&134.40$\pm$0.19&53.73$\pm$0.09&42.25$\pm$0.08\\ \hline
\end{tabular}
\end{table}

%% file: tabintb1_2a.tex
\begin{table}[t]
\caption{continued}
\begin{tabular}{lrrrrrr} \hline
Line             & \multicolumn{1}{c}{$\lambda$}& \multicolumn{5}{c}{100$\times$$I(\lambda)$/$I$(H$\beta$)$^{\rm a}$}\\ \cline{3-7}
& & \multicolumn{5}{c}{Galaxy} \\ \cline{3-7}
 & & J1310$+$0852&J1323$-$0132&J1325$+$0744&J1411$+$0550&J1526$+$2113 \\ \hline
$[$Fe~{\sc iv}$]$    & 3096.00&...~~~~~     &...~~~~~     &...~~~~~     &  1.0$\pm$0.3&...~~~~~     \\
He~{\sc i}           & 3187.74&  2.9$\pm$0.5&  5.0$\pm$0.4&  2.8$\pm$0.6&  3.5$\pm$0.5&  3.2$\pm$0.4\\
He~{\sc ii}          & 3203.10&...~~~~~     &...~~~~~     &...~~~~~     &  0.9$\pm$0.2&  0.8$\pm$0.3\\
O~{\sc iii}          & 3444.07&...~~~~~     &...~~~~~     &...~~~~~     &  0.4$\pm$0.1&...~~~~~     \\
He~{\sc i}           & 3587.28&...~~~~~     &  0.5$\pm$0.1&  0.7$\pm$0.3&  0.2$\pm$0.1&...~~~~~     \\
He~{\sc i}           & 3634.25&  0.8$\pm$0.1&  0.6$\pm$0.1&  0.7$\pm$0.1&  0.6$\pm$0.1&  0.3$\pm$0.1\\
H22                  & 3676.37&  0.5$\pm$0.1&  0.6$\pm$0.1&  0.7$\pm$0.1&  0.6$\pm$0.1&  0.8$\pm$0.1\\
H21                  & 3679.31&  1.4$\pm$0.1&  0.6$\pm$0.1&  0.7$\pm$0.1&  0.7$\pm$0.1&  1.1$\pm$0.1\\
H20                  & 3682.81&  1.0$\pm$0.1&  0.8$\pm$0.1&  0.7$\pm$0.1&  0.8$\pm$0.1&  1.2$\pm$0.1\\
H19                  & 3686.83&  1.0$\pm$0.1&  0.9$\pm$0.1&  1.1$\pm$0.1&  2.0$\pm$0.1&  1.2$\pm$0.1\\
H18                  & 3691.55&  1.0$\pm$0.1&  0.7$\pm$0.1&  1.0$\pm$0.1&  1.3$\pm$0.1&  1.2$\pm$0.1\\
H17                  & 3697.15&  1.6$\pm$0.2&  1.5$\pm$0.1&  1.4$\pm$0.2&  1.3$\pm$0.1&  1.2$\pm$0.1\\
H16+O~{\sc iii}      & 3703.30&  2.4$\pm$0.2&...~~~~~     &  2.7$\pm$0.2&  1.9$\pm$0.1&  2.2$\pm$0.1\\
H15                  & 3711.97&  2.6$\pm$0.2&  2.4$\pm$0.2&  3.4$\pm$0.4&  2.3$\pm$0.1&  2.6$\pm$0.2\\
H14                  & 3721.94&  3.0$\pm$0.2&  2.9$\pm$0.2&  2.9$\pm$0.2&  3.0$\pm$0.1&  2.4$\pm$0.1\\
$[$O~{\sc ii}$]$     & 3726.03& 16.5$\pm$0.6&  8.6$\pm$0.3& 20.8$\pm$0.7& 22.3$\pm$0.7& 11.5$\pm$0.4\\
$[$O~{\sc ii}$]$     & 3728.82& 23.4$\pm$0.8&  9.6$\pm$0.3& 24.2$\pm$0.8& 25.8$\pm$0.8& 13.6$\pm$0.5\\
H13                  & 3734.37&  2.4$\pm$0.1&  2.8$\pm$0.2&  2.2$\pm$0.2&  2.7$\pm$0.1&  2.4$\pm$0.1\\
H12                  & 3750.15&  3.6$\pm$0.2&  3.5$\pm$0.3&  3.9$\pm$0.5&  3.7$\pm$0.1&  4.1$\pm$0.2\\
H11                  & 3770.63&  4.5$\pm$0.2&  4.2$\pm$0.4&  5.1$\pm$0.7&  4.4$\pm$0.2&  4.9$\pm$0.2\\
H10                  & 3797.90&  6.1$\pm$0.3&  5.7$\pm$0.3&  6.3$\pm$0.4&  5.7$\pm$0.2&  6.4$\pm$0.3\\
He~{\sc i}           & 3819.64&  1.0$\pm$0.1&  1.2$\pm$0.1&  1.4$\pm$0.2&  1.0$\pm$0.1&  1.0$\pm$0.2\\
H9                   & 3835.39&  8.4$\pm$0.2&  7.6$\pm$0.3&  8.3$\pm$0.4&  8.0$\pm$0.3&  7.8$\pm$0.3\\
$[$Ne~{\sc iii}$]$   & 3868.76& 55.2$\pm$1.7& 59.1$\pm$1.8& 56.6$\pm$1.8& 54.8$\pm$1.7& 49.1$\pm$1.5\\
He~{\sc i}+H8        & 3889.00& 19.0$\pm$0.6& 20.7$\pm$0.7& 18.5$\pm$0.8& 18.9$\pm$0.6& 18.9$\pm$0.6\\
He~{\sc i}           & 3964.73&  0.5$\pm$0.1&  0.8$\pm$0.1&  1.1$\pm$0.2&  0.6$\pm$0.1&  0.5$\pm$0.1\\
$[$Ne~{\sc iii}$]$   & 3967.46& 18.2$\pm$0.6& 18.5$\pm$0.6& 18.2$\pm$0.7& 17.2$\pm$0.5& 17.1$\pm$0.6\\
H7                   & 3970.07& 14.8$\pm$0.5& 16.1$\pm$0.5& 14.8$\pm$0.5& 15.6$\pm$0.5& 14.2$\pm$0.5\\
He~{\sc i}           & 4026.19&  1.8$\pm$0.1&  2.2$\pm$0.1&  1.7$\pm$0.1&  0.3$\pm$0.1&  1.8$\pm$0.2\\
$[$S~{\sc ii}$]$     & 4068.60&  0.8$\pm$0.2&  0.3$\pm$0.1&  0.8$\pm$0.1&  0.9$\pm$0.1&  0.5$\pm$0.1\\
$[$S~{\sc ii}$]$     & 4076.35&...~~~~~     &...~~~~~     &...~~~~~     &  0.3$\pm$0.1&...~~~~~     \\
H$\delta$            & 4101.74& 23.7$\pm$0.7& 26.9$\pm$0.8& 26.1$\pm$0.9& 26.4$\pm$0.8& 25.4$\pm$0.8\\
He~{\sc i}           & 4120.84&...~~~~~     &  0.4$\pm$0.1&...~~~~~     &  0.2$\pm$0.1&  0.3$\pm$0.1\\
He~{\sc i}           & 4143.76&...~~~~~     &  0.5$\pm$0.1&...~~~~~     &  0.3$\pm$0.1&  0.4$\pm$0.1\\
$[$Fe~{\sc v}$]$     & 4227.20&...~~~~~     &  0.7$\pm$0.1&...~~~~~     &  0.3$\pm$0.1&  0.5$\pm$0.1\\
$[$Fe~{\sc ii}$]$    & 4287.33&...~~~~~     &...~~~~~     &...~~~~~     &  0.4$\pm$0.1&...~~~~~     \\
H$\gamma$            & 4340.47& 45.7$\pm$1.4& 48.4$\pm$1.4& 46.9$\pm$1.4& 47.8$\pm$1.4& 43.6$\pm$1.3\\
$[$O~{\sc iii}$]$    & 4363.21& 17.1$\pm$0.5& 21.3$\pm$0.6& 13.9$\pm$0.4& 15.4$\pm$0.5& 18.5$\pm$0.6\\
He~{\sc i}           & 4387.93&  0.6$\pm$0.1&  0.6$\pm$0.1&  0.5$\pm$0.1&  0.4$\pm$0.1&  0.5$\pm$0.1\\
He~{\sc i}           & 4471.48&  3.9$\pm$0.1&  4.0$\pm$0.1&  4.1$\pm$0.3&  4.0$\pm$0.1&  4.1$\pm$0.1\\
$[$Fe~{\sc iii}$]$   & 4658.10&  0.4$\pm$0.1&  0.5$\pm$0.1&  0.8$\pm$0.1&  0.8$\pm$0.1&  0.3$\pm$0.1\\
He~{\sc ii}          & 4686.00&  0.6$\pm$0.1&  1.8$\pm$0.1&  0.4$\pm$0.1&  1.7$\pm$0.1&  1.4$\pm$0.1\\
$[$Ar~{\sc iv}$]$    & 4711.37&  3.1$\pm$0.1&  4.1$\pm$0.2&  2.5$\pm$0.1&  2.1$\pm$0.1&  2.7$\pm$0.1\\
He~{\sc i}           & 4713.14&  0.5$\pm$0.1&  0.7$\pm$0.1&  0.5$\pm$0.1&  0.7$\pm$0.1&  0.6$\pm$0.1\\
$[$Ar~{\sc iv}$]$    & 4740.20&  2.5$\pm$0.1&  3.4$\pm$0.1&  2.4$\pm$0.1&  1.8$\pm$0.1&  2.3$\pm$0.1\\
H$\beta$             & 4861.33&100.0$\pm$2.9&100.0$\pm$2.8&100.0$\pm$2.9&100.0$\pm$2.9&100.0$\pm$2.9\\
$[$Fe~{\sc ii}$]$    & 4905.00&...~~~~~     &  0.2$\pm$0.1&...~~~~~     &  0.4$\pm$0.1&...~~~~~     \\
He~{\sc i}           & 4921.93&  1.1$\pm$0.1&  1.2$\pm$0.1&  1.7$\pm$0.1&  1.2$\pm$0.1&  1.3$\pm$0.1\\
$[$Fe~{\sc iii}$]$   & 4930.50& ...~~~~~    &  0.3$\pm$0.1&...~~~~~     &...~~~~~     &...~~~~~     \\
$[$O~{\sc iii}$]$    & 4958.92&256.8$\pm$7.4&245.5$\pm$7.0&255.5$\pm$7.3&243.3$\pm$6.9&232.3$\pm$6.7\\
$[$Fe~{\sc iii}$]$   & 4988.00&  0.8$\pm$0.1&  0.3$\pm$0.1&  0.6$\pm$0.1&  0.8$\pm$0.1&  0.5$\pm$0.1\\
$[$O~{\sc iii}$]$    & 5006.85&769.2$\pm$22.&724.7$\pm$21.&773.5$\pm$22.&729.5$\pm$21.&701.0$\pm$20.\\
He~{\sc i}           & 5015.68&  2.0$\pm$0.1&  2.0$\pm$0.1&  2.2$\pm$0.1&  1.8$\pm$0.1&  1.7$\pm$0.1\\
Si~{\sc ii}          & 5041.03&...~~~~~     &...~~~~~     &...~~~~~     &  0.3$\pm$0.1&...~~~~~     \\
He~{\sc i}           & 5047.80&...~~~~~     &  0.3$\pm$0.1&...~~~~~     &  0.2$\pm$0.1&  0.3$\pm$0.1\\
$[$Cl~{\sc iii}$]$   & 5517.71&...~~~~~     &...~~~~~     &...~~~~~     &  0.4$\pm$0.1&  0.4$\pm$0.1\\
$[$Cl~{\sc iii}$]$   & 5537.88&...~~~~~     &...~~~~~     &...~~~~~     &  0.3$\pm$0.1&...~~~~~     \\
$[$N~{\sc ii}$]$     & 5754.44&...~~~~~     &...~~~~~     &...~~~~~     &  0.2$\pm$0.1&...~~~~~     \\ \hline
\end{tabular}
\end{table}

%% file: tabintb1_2b.tex
\begin{table}[t]
\caption{continued}
\begin{tabular}{lrrrrrr} \hline
Line             & \multicolumn{1}{c}{$\lambda$}& \multicolumn{5}{c}{100$\times$$I(\lambda)$/$I$(H$\beta$)$^{\rm a}$}\\ \cline{3-7}
& & \multicolumn{5}{c}{Galaxy} \\ \cline{3-7}
 & & J1310$+$0852&J1323$-$0132&J1325$+$0744&J1411$+$0550&J1526$+$2113 \\ \hline
He~{\sc i}           & 5875.60& 10.9$\pm$0.4& 10.5$\pm$0.3& 11.6$\pm$0.4& 11.4$\pm$0.4& 11.1$\pm$0.4\\
$[$O~{\sc i}$]$      & 6300.30&  1.0$\pm$0.1&  0.5$\pm$0.1&  1.6$\pm$0.1&  1.5$\pm$0.1&  0.9$\pm$0.1\\
$[$S~{\sc iii}$]$    & 6312.10&  1.5$\pm$0.1&  0.9$\pm$0.1&  1.8$\pm$0.1&  1.4$\pm$0.1&  0.9$\pm$0.1\\
$[$O~{\sc i}$]$      & 6363.80&  0.3$\pm$0.1&  0.2$\pm$0.1&  0.4$\pm$0.1&  0.5$\pm$0.1&  0.4$\pm$0.1\\
$[$N~{\sc ii}$]$     & 6548.10&  0.7$\pm$0.1&  0.3$\pm$0.1&  1.3$\pm$0.1&  2.0$\pm$0.2&  0.7$\pm$0.1\\
H$\alpha$            & 6562.80&275.7$\pm$8.6&276.5$\pm$8.5&280.5$\pm$8.7&279.7$\pm$8.6&277.7$\pm$8.6\\
$[$N~{\sc ii}$]$     & 6583.40&  2.1$\pm$0.1&  1.1$\pm$0.1&  3.3$\pm$0.1&  5.7$\pm$0.2&  2.3$\pm$0.1\\
He~{\sc i}           & 6678.10&  3.0$\pm$0.1&  2.8$\pm$0.1&  3.3$\pm$0.2&  3.0$\pm$0.1&  3.1$\pm$0.1\\
$[$S~{\sc ii}$]$     & 6716.40&  4.5$\pm$0.2&  1.3$\pm$0.1&  5.1$\pm$0.3&  4.4$\pm$0.1&  2.7$\pm$0.1\\
$[$S~{\sc ii}$]$     & 6730.80&  3.2$\pm$0.1&  1.1$\pm$0.1&  4.3$\pm$0.2&  3.5$\pm$0.1&  2.1$\pm$0.1\\
He~{\sc i}           & 7065.30&  3.3$\pm$0.1&  3.1$\pm$0.2&  4.3$\pm$0.2&  4.7$\pm$0.2&  4.7$\pm$0.4\\
$[$Ar~{\sc iii}$]$   & 7135.80&  5.3$\pm$0.2&  3.0$\pm$0.1&  6.4$\pm$0.2&  4.8$\pm$0.2&  3.1$\pm$0.2\\
$[$Ar~{\sc iv}$]$    & 7262.76&...~~~~~     &  0.2$\pm$0.1&...~~~~~     &...~~~~~     &...~~~~~     \\
He~{\sc i}           & 7281.35&...~~~~~     &  0.5$\pm$0.1&...~~~~~     &  0.5$\pm$0.1&  1.1$\pm$0.1\\
$[$O~{\sc ii}$]$     & 7319.90&  0.9$\pm$0.1&  0.5$\pm$0.1&  0.9$\pm$0.1&  1.2$\pm$0.1&  0.8$\pm$0.1\\
$[$O~{\sc ii}$]$     & 7330.20&  0.7$\pm$0.1&  0.3$\pm$0.1&  0.9$\pm$0.1&  1.0$\pm$0.1&  0.6$\pm$0.1\\
$[$Ar~{\sc iii}$]$   & 7751.12&  1.4$\pm$0.1&  0.8$\pm$0.1&  1.6$\pm$0.2&  1.2$\pm$0.1&  0.9$\pm$0.1\\
$[$Cl~{\sc iv}$]$    & 8045.63&...~~~~~     &  0.6$\pm$0.1&...~~~~~     &  0.2$\pm$0.1&  0.4$\pm$0.1\\
P20                  & 8392.40&...~~~~~     &  0.3$\pm$0.1&...~~~~~     &  0.3$\pm$0.1&  0.2$\pm$0.1\\
P19                  & 8413.32&  0.3$\pm$0.1&  0.3$\pm$0.1&...~~~~~     &  0.4$\pm$0.1&  0.5$\pm$0.1\\
P18                  & 8437.96&  0.2$\pm$0.1&  0.3$\pm$0.1&  0.3$\pm$0.1&  0.4$\pm$0.1&  0.3$\pm$0.1\\
O~{\sc i}            & 8446.34&...~~~~~     &  0.2$\pm$0.1&  0.3$\pm$0.1&  0.5$\pm$0.1&  0.4$\pm$0.1\\
P17                  & 8467.26&  0.5$\pm$0.1&  0.4$\pm$0.1&  0.3$\pm$0.1&  0.3$\pm$0.1&  0.5$\pm$0.1\\
P16                  & 8502.49&  0.4$\pm$0.1&  0.4$\pm$0.1&  0.3$\pm$0.1&  0.5$\pm$0.1&  0.5$\pm$0.1\\
P15                  & 8545.38&  0.5$\pm$0.1&  0.5$\pm$0.1&  0.5$\pm$0.1&  0.6$\pm$0.1&  0.6$\pm$0.1\\
P14                  & 8598.39&  0.8$\pm$0.1&  0.5$\pm$0.1&  0.8$\pm$0.1&  0.8$\pm$0.1&  1.0$\pm$0.1\\
P13                  & 8665.02&  0.9$\pm$0.1&  0.8$\pm$0.1&...~~~~~     &  1.1$\pm$0.1&  1.0$\pm$0.1\\
P12                  & 8750.47&  1.1$\pm$0.1&  1.0$\pm$0.1&  1.1$\pm$0.1&  1.2$\pm$0.2&  1.0$\pm$0.1\\
P11                  & 8862.79&  0.9$\pm$0.1&  1.3$\pm$0.1&...~~~~~     &  1.5$\pm$0.1&  1.3$\pm$0.1\\
P10                  & 9014.91&  2.2$\pm$0.1&  1.5$\pm$0.2&  1.9$\pm$0.2&  1.1$\pm$0.1&  2.3$\pm$0.1\\
$[$S~{\sc iii}$]$    & 9069.00&  7.4$\pm$0.4&  4.2$\pm$0.2& 13.6$\pm$0.6&  7.0$\pm$0.3&  6.0$\pm$0.3\\
P9                   & 9229.02&  2.6$\pm$0.2&  2.3$\pm$0.1&...~~~~~     &  2.3$\pm$0.1&  2.2$\pm$0.4\\
$[$S~{\sc iii}$]$    & 9530.60& 23.1$\pm$1.0& 11.9$\pm$0.6& 36.8$\pm$1.6& 22.1$\pm$0.8& 13.7$\pm$0.7\\
P8                   & 9545.98&  4.1$\pm$0.4&  2.9$\pm$0.3&  5.7$\pm$0.6&  3.7$\pm$0.2&  3.7$\pm$0.4\\
P$\delta$            &10052.15&  6.3$\pm$0.3&  4.8$\pm$0.3&  6.2$\pm$0.4&  5.2$\pm$0.2&  5.2$\pm$0.2\\
He~{\sc i}           &10830.00&  5.4$\pm$0.3& 31.1$\pm$1.2& 52.5$\pm$2.2& 53.1$\pm$2.1& 35.8$\pm$1.4\\
P$\gamma$            &10941.12&  8.0$\pm$0.4&  7.3$\pm$0.4&  9.1$\pm$0.5&  8.9$\pm$0.4&  9.1$\pm$0.4\\
He~{\sc i}           &12790.00&...~~~~~     &  0.5$\pm$0.1&...~~~~~     &  1.4$\pm$0.1&...~~~~~     \\
P$\beta$             &12821.62& 15.2$\pm$0.7& 14.1$\pm$0.6&  6.1$\pm$0.4& 11.9$\pm$0.5&  0.8$\pm$0.1\\
Br15                 &15704.96&...~~~~~     &  0.3$\pm$0.1&...~~~~~     &  0.3$\pm$0.1&  0.3$\pm$0.1\\
Br14                 &15884.88&...~~~~~     &  0.3$\pm$0.1&...~~~~~     &...~~~~~     &  0.3$\pm$0.1\\
Br13                 &16113.92&...~~~~~     &...~~~~~     &...~~~~~     &  0.5$\pm$0.1&  0.4$\pm$0.1\\
Br12                 &16411.68&  0.7$\pm$0.1&  0.6$\pm$0.1&...~~~~~     &  0.7$\pm$0.1&  0.4$\pm$0.1\\
Br11                 &16811.12&  0.8$\pm$0.1&  0.5$\pm$0.1&...~~~~~     &  1.2$\pm$0.1&...~~~~~     \\
Br10                 &17366.86&...~~~~~     &  0.7$\pm$0.1&...~~~~~     &...~~~~~     &...~~~~~     \\
Br9                  &18179.09&...~~~~~     &  0.9$\pm$0.1&...~~~~~     &  0.7$\pm$0.1&  1.1$\pm$0.1\\
P$\alpha$            &18756.20&...~~~~~     & 27.6$\pm$1.2& 20.5$\pm$0.9& 28.3$\pm$1.2& 20.2$\pm$0.9\\
Br$\delta$           &19450.88&  1.9$\pm$0.1&  1.8$\pm$0.1&  2.4$\pm$0.2&  0.7$\pm$0.1&  1.7$\pm$0.1\\
He~{\sc i}           &20580.00&  0.9$\pm$0.1&  0.5$\pm$0.1&...~~~~~     &  1.4$\pm$0.1&...~~~~~     \\
Br$\gamma$           &21661.21&  3.3$\pm$0.1&  1.9$\pm$0.1&...~~~~~     &  2.9$\pm$0.1&...~~~~~     \\ \\
$C$(H$\beta$)$^{\rm b}$        &        &0.000$\pm$0.037&0.115$\pm$0.037&0.040$\pm$0.037&0.095$\pm$0.037&0.120$\pm$0.037\\
$EW$(H$\beta$)$^{\rm c}$       &        &271.5$\pm$0.6&298.1$\pm$0.5&346.2$\pm$1.0&265.5$\pm$0.3&389.5$\pm$1.0\\
$F$(H$\beta$)$^{\rm d}$        &        &28.51$\pm$0.07&101.8$\pm$0.2&20.46$\pm$0.06&55.79$\pm$0.07&35.70$\pm$0.09\\ \hline
\end{tabular}
\tablefoot{$^{\rm a}$$I(\lambda)$, $I$(H$\beta$) are extinction-corrected intensities. $^{\rm b}$Extinction coefficient. $^{\rm c}$Rest-frame equivalent width of the H$\beta$ emission line. $^{\rm d}$Observed intensity of the H$\beta$ emission line in 10$^{-16}$ erg cm$^{-2}$ s$^{-1}$.}
\end{table}

%% file: tabcheb3_1a.tex
\begin{table}
  \caption{Electron temperatures, electron number densities and element
    abundances} \label{taba2}
\begin{tabular}{lcccccc} \hline
Property                             &\multicolumn{6}{c}{Value} \\ \cline{2-7}
& J0007$+$0226&J0009$+$0234&J0159$+$0751&J0240$-$0828&J0825$+$1846&J1255$-$0213 \\ \hline
$T_{\rm e}$(O {\sc iii}), K          &17660$\pm$360&19170$\pm$440&20120$\pm$460&15220$\pm$260&17950$\pm$380&16530$\pm$310       \\
$T_{\rm e}$(O {\sc ii}), K           &15170$\pm$290&15510$\pm$330&16000$\pm$330&14120$\pm$230&15250$\pm$300&14760$\pm$260\\
$T_{\rm e}$(S {\sc iii}), K          &16130$\pm$300&17980$\pm$370&18410$\pm$380&14000$\pm$220&17130$\pm$320&15490$\pm$260       \\
$N_{\rm e}$(S {\sc ii}), cm$^{-3}$    &10$\pm$10&125$\pm$83&170$\pm$110&870$\pm$220&28$\pm$46&100$\pm$68       \\
$N_{\rm e}$(O {\sc ii}), cm$^{-3}$    &148$\pm$69&49$\pm$55&199$\pm$68&574$\pm$81&182$\pm$57&68$\pm$61        \\ \\
\underline{Helium (5 lines):} \\
$T_{\rm e}$(He), K &16450&18200&18430&15650&18620&16810 \\
$N_{\rm e}$(He), cm$^{-3}$  &12&11&821& 721  &32&183       \\
$\tau$($\lambda$3889)&3.7&0.03&4.97&3.86&0.84&0.02 \\
Y(mean)    &0.2647&0.2808&0.2326&0.2550   &0.2811&0.2591       \\
\underline{Helium (6 lines):} \\
$T_{\rm e}$(He), K &16360&18190&20270&15660&18600&16840 \\
$N_{\rm e}$(He), cm$^{-3}$  &143&10&90&570&10&25       \\
$\tau$($\lambda$3889)&3.36&0.12&5.00&4.54&1.11&0.39 \\
Y(mean)   &0.2566&0.2660&0.2588&0.2606&0.2798&0.2686       \\ \\
O$^+$/H$^+$$\times$10$^6$            &1.803$\pm$0.130&3.488$\pm$0.253&1.173$\pm$0.083&5.490$\pm$0.353&4.173$\pm$0.286&5.716$\pm$0.372 \\
O$^{2+}$/H$^+$$\times$10$^5$          &5.933$\pm$0.315&3.022$\pm$0.174&3.634$\pm$0.204&7.661$\pm$0.383&3.765$\pm$0.208&5.482$\pm$0.283 \\
O$^{3+}$/H$^+$$\times$10$^6$          &1.060$\pm$0.099&0.557$\pm$0.058&0.549$\pm$0.045&2.362$\pm$0.229&1.071$\pm$0.094&1.241$\pm$0.088\\
12+log(O/H)                         &7.794$\pm$0.022&7.535$\pm$0.022&7.581$\pm$0.023&7.927$\pm$0.020&7.632$\pm$0.021&7.791$\pm$0.020     \\ \\
N$^+$/H$^+$$\times$10$^7$            &1.011$\pm$0.059&1.711$\pm$0.083&0.795$\pm$0.058&7.156$\pm$0.282&1.582$\pm$0.076&2.039$\pm$0.089 \\
ICF(N)                          &29.207&9.003&28.341&13.195&9.374&9.698 \\
log(N/O)                        &$-$1.324$\pm$0.037~~~&$-$1.347$\pm$0.032~~~&$-$1.228$\pm$0.043~~~&$-$0.952$\pm$0.028~~~&$-$1.461$\pm$0.031~~~&$-$1.495$\pm$0.029~~~\\ \\
Ne$^{2+}$/H$^+$$\times$10$^5$        &1.062$\pm$0.060 &0.514$\pm$0.031&0.661$\pm$0.038&1.199$\pm$0.066&0.636$\pm$0.037&1.114$\pm$0.062\\
ICF(Ne)                             &1.002 &1.049&1.018&1.021&1.050&1.041\\
log(Ne/O)                           &$-$0.767$\pm$0.034~~~&$-$0.803$\pm$0.034~~~&$-$0.753$\pm$0.034~~~&$-$0.839$\pm$0.033~~~&$-$0.808$\pm$0.033~~~&$-$0.727$\pm$0.033~~~\\ \\
S$^+$/H$^+$$\times$10$^7$            &0.386$\pm$0.011&0.747$\pm$0.020&0.237$\pm$0.080&1.059$\pm$0.033&0.777$\pm$0.021&0.964$\pm$0.022 \\
S$^{2+}$/H$^+$$\times$10$^7$$^{\rm a}$ &4.587$\pm$0.416&4.393$\pm$0.263&1.203$\pm$0.289&6.829$\pm$0.550&5.313$\pm$0.250&7.342$\pm$0.362 \\
ICF(S)                              &3.540&1.620&3.075&2.250&1.520&1.672  \\
log(S/O)                            &$-$1.548$\pm$0.043~~~&$-$1.614$\pm$0.032~~~&$-$1.934$\pm$0.090~~~&$-$1.677$\pm$0.036~~~&$-$1.666$\pm$0.028~~~&$-$1.666$\pm$0.028~~~\\ \\
S$^{2+}$/H$^+$$\times$10$^7$$^{\rm b}$ &3.422$\pm$0.370&2.117$\pm$0.232&1.144$\pm$0.155&4.234$\pm$0.510&3.651$\pm$0.391&6.661$\pm$0.734 \\
log(S/O)                            &$-$1.664$\pm$0.048~~~&$-$1.868$\pm$0.047~~~&$-$1.868$\pm$0.047~~~&$-$1.851$\pm$0.046~~~&$-$1.805$\pm$0.044~~~&$-$1.686$\pm$0.046~~~\\ \\
S$^{2+}$/H$^+$$\times$10$^7$$^{\rm c}$&3.675$\pm$0.403&2.147$\pm$0.236&1.227$\pm$0.138&4.172$\pm$0.468&4.555$\pm$0.488&5.746$\pm$0.636\\
log(S/O)                            &$-$1.636$\pm$0.048~~~&$-$1.864$\pm$0.042~~~&$-$1.927$\pm$0.047~~~&$-$1.856$\pm$0.044~~~&$-$1.724$\pm$0.045~~~&$-$1.741$\pm$0.046~~~\\ \\
Cl$^{2+}$/H$^+$$\times$10$^8$        &      ... &      ...&...&...&...&1.485$\pm$0.212 \\
ICF(Cl)                             & ...    &      ... &...&...&...&1.500\\
log(Cl/O)                           &  ...  &      ... &...&...&...&$-$3.443$\pm$0.065~~\\ \\
Ar$^{2+}$/H$^+$$\times$10$^7$        &1.463$\pm$0.070&1.246$\pm$0.051 &0.575$\pm$0.039&2.135$\pm$0.093&1.402$\pm$0.055&2.001$\pm$0.079\\
Ar$^{3+}$/H$^+$$\times$10$^7$        &2.262$\pm$0.125&1.104$\pm$0.071 &1.061$\pm$0.078&1.125$\pm$0.073&1.119$\pm$0.045&1.263$\pm$0.064\\
ICF(Ar)                             &2.530&1.364&2.497&1.564&1.381&1.384 \\
log(Ar/O)                           &$-$2.225$\pm$0.048~~~&$-$2.230$\pm$0.038~~~&$-$2.423$\pm$0.070~~~&$-$2.403$\pm$0.031~~~&$-$2.345$\pm$0.034~~~&$-$2.349$\pm$0.030~~~\\ \\
Fe$^{2+}$/H$^+$$\times$10$^7$$^{\rm d}$&      ... &      ...&0.513$\pm$0.072&1.545$\pm$0.123&1.301$\pm$0.165&1.082$\pm$0.092 \\
ICF(Fe)                           & ...    &      ... &44.168&19.591&13.782&14.215\\
log(Fe/O)                           &  ...  &      ... &$-$1.226$\pm$0.066~~~&$-$1.446$\pm$0.040~~~&$-$1.379$\pm$0.059~~~&$-$1.604$\pm$0.042~~~\\ \\
Fe$^{2+}$/H$^+$$\times$10$^7$$^{\rm e}$&      ... &1.131$\pm$0.173&      ... &1.015$\pm$0.084&1.385$\pm$0.152&1.663$\pm$0.103\\
ICF(Fe)                             & ...    &13.264 &      ...&19.591&13.782&14.215 \\
log(Fe/O)                           &  ...  &$-$1.359$\pm$0.070~~~&      ... &$-$1.628$\pm$0.041~~~&$-$1.229$\pm$0.048~~~&$-$1.136$\pm$0.027~~~\\ \hline
\end{tabular}
\end{table}

%% file: tabcheb3_2a.tex
\begin{table}
\caption{continued}
\begin{tabular}{lccccc} \hline
Property                             &\multicolumn{5}{c}{Value} \\ \cline{2-6}
& J1310$+$0852&J1323$-$0132&J1325$+$0744&J1411$+$0550&J1526$+$2113\\
\hline
$T_{\rm e}$(O {\sc iii}), K          &16030$\pm$300&18420$\pm$380&14550$\pm$250&15630$\pm$270&17480$\pm$350\\
$T_{\rm e}$(O {\sc ii}), K           &14530$\pm$250&15370$\pm$300&13730$\pm$220&14340$\pm$240&15110$\pm$280\\
$T_{\rm e}$(S {\sc iii}), K          &14660$\pm$250&16990$\pm$320&13300$\pm$210&14370$\pm$230&16240$\pm$290\\
$N_{\rm e}$(S {\sc ii}), cm$^{-3}$    &10$\pm$10&230$\pm$185&248$\pm$128&147$\pm$69&125$\pm$85\\
$N_{\rm e}$(O {\sc ii}), cm$^{-3}$    &37$\pm$53&297$\pm$61&227$\pm$59&238$\pm$58&222$\pm$61        \\ \\
\underline{Helium (5 lines):} \\
$T_{\rm e}$(He), K&16850&18370&13630&14530&17810 \\
$N_{\rm e}$(He), cm$^{-3}$&163&11&406&679&253       \\
$\tau$($\lambda$3889)&0.73&1.70&4.24&3.45&4.08 \\
Y(mean)    &0.2570&0.2703&0.2571&0.2473&0.2595    \\
\underline{Helium (6 lines):} \\
$T_{\rm e}$(He), K&15270&17810&13620&16000&17810 \\
$N_{\rm e}$(He), cm$^{-3}$&11&142&609&525&186      \\
$\tau$($\lambda$3889)&4.94&1.25&3.54&3.80&4.15 \\
Y(mean)   &0.1314&0.2603&0.2508&0.2488&0.2634      \\ \\
O$^+$/H$^+$$\times$10$^6$            &3.997$\pm$0.261&1.571$\pm$0.110&5.572$\pm$0.364&5.112$\pm$0.323&2.249$\pm$0.151\\
O$^{2+}$/H$^+$$\times$10$^5$          &7.188$\pm$0.369&4.947$\pm$0.265&9.175$\pm$0.462&7.245$\pm$0.361&5.343$\pm$0.282\\
O$^{3+}$/H$^+$$\times$10$^6$          &0.434$\pm$0.053&0.974$\pm$0.083&0.364$\pm$0.064&1.353$\pm$0.093&0.761$\pm$0.061\\
12+log(O/H)                         &7.883$\pm$0.021&7.716$\pm$0.022&7.990$\pm$0.021&7.897$\pm$0.020&7.752$\pm$0.022\\ \\
N$^+$/H$^+$$\times$10$^7$            &1.629$\pm$0.088&0.789$\pm$0.055&2.948$\pm$0.128&4.620$\pm$0.168&1.864$\pm$0.088\\
ICF(N)                              &16.283&28.450&14.690&13.306&21.624\\
log(N/O)                            &$-$1.459$\pm$0.035~~~&$-$1.365$\pm$0.042~~~&$-$1.353$\pm$0.031~~~&$-$1.108$\pm$0.028~~~&$-$1.190$\pm$0.034~~~\\ \\
Ne$^{2+}$/H$^+$$\times$10$^5$        &1.224$\pm$0.069&0.911$\pm$0.051&1.658$\pm$0.093&1.304$\pm$0.071&0.863$\pm$0.048\\
ICF(Ne)                             &1.000&1.011&0.990&1.053&1.010\\
log(Ne/O)                           &$-$0.795$\pm$0.033~~~&$-$0.752$\pm$0.034~~~&$-$0.775$\pm$0.033~~~&$-$0.775$\pm$0.032~~~&$-$0.811$\pm$0.034~~~\\ \\
S$^+$/H$^+$$\times$10$^7$            &0.775$\pm$0.021&0.223$\pm$0.011&1.091$\pm$0.039&0.831$\pm$0.019&0.456$\pm$0.013\\
S$^{2+}$/H$^+$$\times$10$^6$$^{\rm a}$ &0.814$\pm$0.050&0.323$\pm$0.026&1.393$\pm$0.089&0.802$\pm$0.064&0.369$\pm$0.037\\
ICF(S)                              &2.502&3.192&2.607&2.201&2.698\\
log(S/O)                            &$-$1.534$\pm$0.032~~~&$-$1.673$\pm$0.039~~~&$-$1.397$\pm$0.033~~~&$-$1.608$\pm$0.037~~~&$-$1.703$\pm$0.044~~~\\ \\
S$^{2+}$/H$^+$$\times$10$^6$$^{\rm b}$ &0.464$\pm$0.053&0.218$\pm$0.024&0.988$\pm$0.111&0.456$\pm$0.050&0.329$\pm$0.036\\
log(S/O)                            &$-$1.751$\pm$0.047~~~&$-$1.831$\pm$0.048~~~&$-$1.533$\pm$0.049~~~&$-$1.823$\pm$0.045~~~&$-$1.748$\pm$0.047~~~\\ \\
S$^{2+}$/H$^+$$\times$10$^6$$^{\rm c}$&0.581$\pm$0.6063&0.249$\pm$0.027&1.078$\pm$0.121&0.579$\pm$0.063&0.305$\pm$0.034\\
log(S/O)                            &$-$1.666$\pm$0.047~~~&$-$1.779$\pm$0.049~~~&$-$1.499$\pm$0.049~~~&$-$1.734$\pm$0.046~~~&$-$1.776$\pm$0.048~~~\\ \\
Cl$^{2+}$/H$^+$$\times$10$^8$        &      ...&      ...&      ... &1.727$\pm$0.212&1.547$\pm$0.060 \\
ICF(Cl)                             & ...   &      ... &      ... &1.532&1.540\\
log(Cl/O)                           &  ...  &      ...&      ... &$-$3.475$\pm$0.057~~~&$-$3.375$\pm$0.028~~~\\ \\
Ar$^{2+}$/H$^+$$\times$10$^7$        &2.218$\pm$0.093&0.989$\pm$0.043&3.203$\pm$0.143&2.067$\pm$0.081&1.096$\pm$0.062\\
Ar$^{3+}$/H$^+$$\times$10$^7$        &2.098$\pm$0.109&2.058$\pm$0.100&2.650$\pm$0.181&1.639$\pm$0.076&1.590$\pm$0.081\\
ICF(Ar)                             &1.751&2.493&1.639&1.575&2.083\\
log(Ar/O)                           &$-$2.293$\pm$0.035~~~&$-$2.324$\pm$0.053~~~&$-$2.270$\pm$0.037~~~&$-$2.385$\pm$0.031~~~&$-$2.389$\pm$0.046~~~\\ \\
Fe$^{2+}$/H$^+$$\times$10$^7$$^{\rm d}$&0.852$\pm$0.149&0.829$\pm$0.168&0.187$\pm$0.019&1.657$\pm$0.096&0.503$\pm$0.078\\
ICF(Fe)                             &24.512&44.111&21.834&19.803&33.175\\
log(Fe/O)                            &$-$1.563$\pm$0.079~~~&$-$1.153$\pm$0.091~~~&$-$1.379$\pm$0.050~~~&$-$1.381$\pm$0.032~~~&$-$1.530$\pm$0.071~~~\\ \\
Fe$^{2+}$/H$^+$$\times$10$^7$$^{\rm e}$&1.540$\pm$0.148&0.520$\pm$0.107&1.379$\pm$0.185&1.624$\pm$0.102&1.008$\pm$0.104\\
ICF(Fe)                             &24.512&44.111&21.834&19.803&33.175\\
log(Fe/O)                           &$-$1.306$\pm$0.047~~~&$-$1.356$\pm$0.092~~~&$-$1.511$\pm$0.062~~~&$-$1.390$\pm$0.034~~~&$-$1.227$\pm$0.050~~~\\ \hline
\end{tabular}
\tablefoot{$^{\rm a}$S$^{2+}$/H$^+$ is derived using [S~{\sc iii}] $\lambda$6312\AA\ emission line. $^{\rm b}$S$^{2+}$/H$^+$ is derived using [S~{\sc iii}] $\lambda$9069\AA\ emission line. $^{\rm c}$S$^{2+}$/H$^+$ is derived using [S~{\sc iii}] $\lambda$9531\AA\ emission line. $^{\rm d}$Fe$^{2+}$/H$^+$ is derived using [Fe~{\sc iii}] $\lambda$4658\AA\ emission line. $^{\rm e}$Fe$^{2+}$/H$^+$ is derived using [Fe~{\sc iii}] $\lambda$4988\AA\ emission line.}
\end{table}